\documentclass[journal]{IEEEtran}
\pdfoutput=1
\usepackage{graphicx}
\usepackage{amssymb}
\usepackage{color}

\begin{document}
\title{Breaking Bandwidth Limit: A Review of Broadband Doherty Power Amplifier Design for 5G}

\author{Gholamreza~Nikandish,~\IEEEmembership{Member,~IEEE,}~Robert~Bogdan~Staszewski,~\IEEEmembership{Fellow,~IEEE,}~\\and~Anding~Zhu,~\IEEEmembership{Senior Member,~IEEE}
\thanks{
This paper has been accepted for publication in IEEE Microwave Magazine. \par
\textcopyright~2019 IEEE. Personal use of this material is permitted. Permission from IEEE must be obtained for all other uses, in any current or future media, including reprinting/republishing this material for advertising or promotional purposes, creating new collective works, for resale or redistribution to servers or lists, or reuse of any copyrighted component of this work in other works. \par
The authors are with the School of Electrical and Electronic Engineering, University College Dublin, Ireland (e-mail: nikandish@ucd.ie, robert.staszewski@ucd.ie, anding.zhu@ucd.ie).}}

\maketitle

\IEEEpeerreviewmaketitle

\section{Introduction}

\IEEEPARstart{T}{he} next generation wireless network, 5G, is expected to provide ubiquitous connections to billions of devices as well as to unlock many new services with multi-Gigabit-per-second data transmission. To meet ever increasing demands for higher data rates and larger capacity, new modulation schemes have been developed and wider frequency bands, e.g., at mm-wave, have been designated to 5G \cite{rappaport13}, \cite{NGMN15}. Massive multi-input multi-output (MIMO), that uses a large number of antennas at the transmitter and receiver, has been considered as one of the key enabling technologies in 5G to improve data throughput and spectrum efficiency \cite{puglielli16}. These new application scenarios pose stringent requirements on the wireless transceiver front-ends and call for special considerations at both circuit and system design levels. In the transmitter, power amplifiers (PAs) should accommodate complex modulated signals, featuring high peak-to-average power ratio (PAPR) and wide modulation bandwidth. Moreover, in massive MIMO arrays, the PAs should maintain high average efficiency to mitigate thermal cooling issues.

Several PA architectures have been adopted to efficiently amplify signals with large PAPR. The popular architectures include envelope tracking, outphasing and Doherty. Since its first introduction in 1936 \cite{doherty36}, the Doherty power amplifier (DPA) has been extensively explored and it has become one of the most widely used PA architectures in the existing cellular base stations. The DPA basically consists of two amplifiers having their output power combined through a load modulation network. It can maintain high efficiency over a large power range and it features a simple circuit implementation compared to the other architectures. Recent research also shows that it has a capability of operating at mm-wave frequencies \cite{asbeck16}. Unfortunately, the classical DPA suffers intrinsic bandwidth limitations, mainly due to narrow-band quarter-wavelength transmissions lines used for impedance transformation. Bandwidth extension is thus an important consideration in modern DPA designs and it has received increasing attention in recent research, especially for wideband 5G applications. There are number of review papers on DPAs published in the past years \cite{grebennikov12}--\cite{pengelly16}. However, there is no complete review on broadband design techniques for the DPA, an essential subject for 5G wireless transmitters.

In this paper, we present a comprehensive review and critical discussion on bandwidth enhancement techniques for the DPA proposed in the literature, in order to provide a thorough understating of broadband design of DPA for high-efficiency 5G wireless transmitters. The paper is organized as follows. In Section II, we discuss the main bandwidth limitation factors. In Section III, we review various bandwidth enhancement techniques for the DPA, including modified load modulation networks, frequency response optimization, parasitic compensation, post-matching, distributed DPA, and dual-input digital DPA. This section also covers transformer-based power combining PA and transformer-less load modulated PA architectures, which have a similar operation as in the conventional DPAs. Challenges and design techniques for integrated circuit (IC) implementation of broadband DPA are discussed in Section IV. Finally, concluding remarks are given in Section V.

\section{DPA Bandwidth Limitation}

\begin{figure}[!t]
\centering
\includegraphics[width=3.3in]{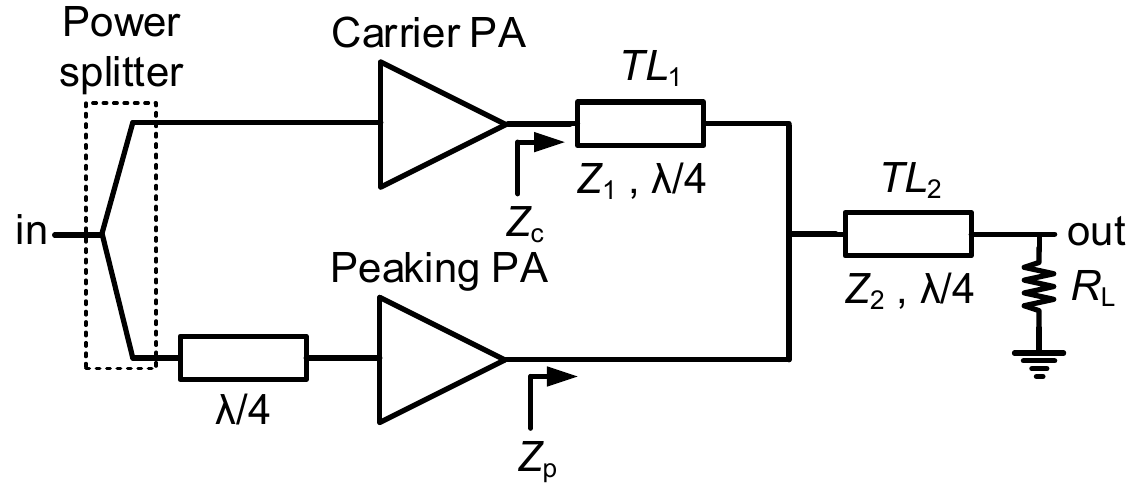}
\caption{The basic DPA architecture.}
\label{DPA-basic}
\end{figure}

The basic DPA architecture is shown in Fig.~\ref{DPA-basic}, where the carrier amplifier is biased in a class-B mode while the peaking amplifier is biased in a class-C mode. At high input power levels, both amplifiers are active and deliver power to the load impedance. The characteristic impedances of the quarter wavelength ($\lambda/4$) transmission lines TL$_1$ and TL$_2$ are chosen such that both amplifiers see their optimum load resistance to provide maximum power and efficiency. When the input power level is low, only the carrier amplifier is active in providing the output power. The load resistance presented to the carrier amplifier is increased by the transmission line TL$_1$, operating as an impedance inverter and serving to improve the DPA efficiency at lower output power levels. The $\lambda/4$ transmission line at the input of peaking amplifier provides a $-90^{\circ}$ phase shift to ensure proper combining of output power from two amplifiers at the common output node. There are several factors contributing to bandwidth limitation of the DPA which will be discussed in the following. We provide new insights using theoretical derivation of the impedance presented to the carrier amplifier at the peak power and back-off.

\begin{figure}[!t]
\centering
\includegraphics[width=2.8in]{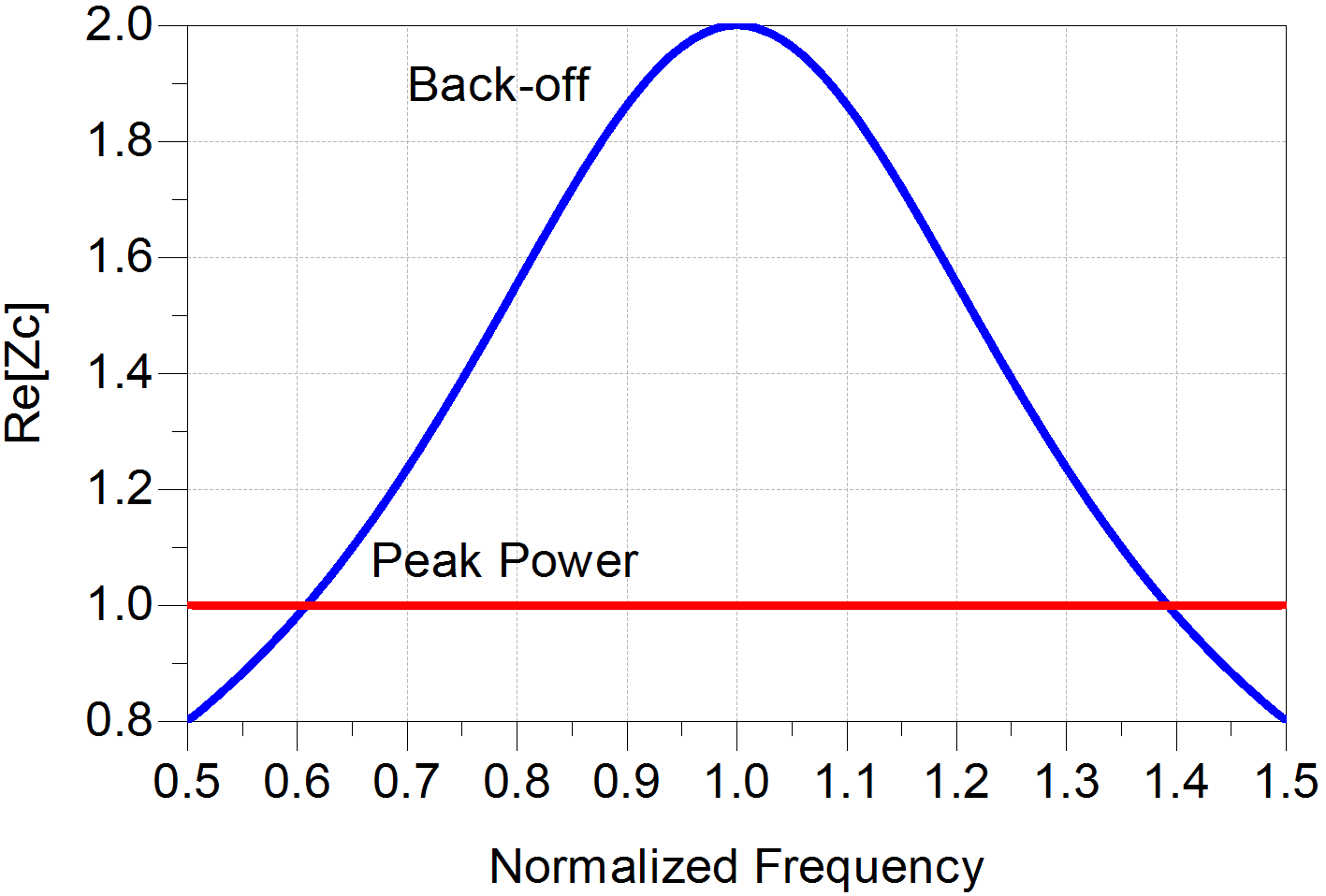}
\caption{Real part of the normalized impedance presented to the carrier amplifier at peak power and 6-dB back-off. The DPA bandwidth is mainly limited by the back-off impedance.}
\label{Zc1}
\end{figure}

\subsection{Output Network}

The main bandwidth-limiting element of the DPA is usually the impedance inverter TL$_1$. This can be shown by exploring the impedances presented to the carrier amplifier at the peak output power and back-off. The characteristic impedances of the $\lambda/4$ transmission lines, assuming a symmetric transistor configuration (i.e., transistors in the carrier and peak amplifiers are identical), are chosen as
\begin{equation}
\label{eq_z1_conv}
Z_1= R_{opt}
\end{equation}
\begin{equation}
\label{eq_z2_conv}
Z_2= \sqrt{\frac{1}{2} R_L R_{opt}},
\end{equation}
where $R_{opt}$ is the optimum load impedance of the carrier and peaking transistors \cite{grebennikov12}. The transmission line TL$_2$ transforms the load impedance $R_L$ into $Z_2^2/R_L=R_{opt}/2$ at the common output node. At peak power, where the two amplifiers deliver identical output currents, each of them sees the optimum impedance, $R_{opt}$. The characteristic impedance of TL$_1$ is chosen to match the optimum load resistance at peak power. Thus, impedance $R_{opt}$ is presented to both amplifiers. At 6-dB back-off, the peaking amplifier is turned off, and the impedance presented to the carrier amplifier, assuming the peaking amplifier presents an open circuit, is $Z_1^2/(R_{opt}/2)=2R_{opt}$. It is noticed that the impedance transformation ratio of the impedance inverter TL$_1$ is 1 at the peak power and 4 at the 6-dB back-off. This limits the bandwidth of the DPA at back-off. In order to illustrate this bandwidth limitation, we compare the impedance presented to the carrier amplifier at the peak power and 6-dB back-off. Assuming that the load presented to the output of impedance inverter TL$_1$ is $kR_{opt}$ ($k=1$ at peak power and $k=0.5$ at back-off), it can be shown that
\begin{equation}
\label{eq_Zc1}
Z_{c}(f) = R_{opt} \frac{k + j \tan (\frac{\pi}{2} \frac{f}{f_0})}{1+jk \tan (\frac{\pi}{2} \frac{f}{f_0})}.
\end{equation}

The real part of $Z_{c}(f)$, normalized to $R_{opt}$, at the peak power and 6-dB back-off is depicted in Fig.~\ref{Zc1}. A narrower bandwidth is observed at back-off. The fractional bandwidth for 20\% reduction in the real part of the impedance, roughly corresponding to 1-dB drop in output power ($P_{out}=Re\{Z_L\}I_m^2/2$), is 38\% at back-off.

\begin{figure}[!t]
\centering
\includegraphics[width=2.8in]{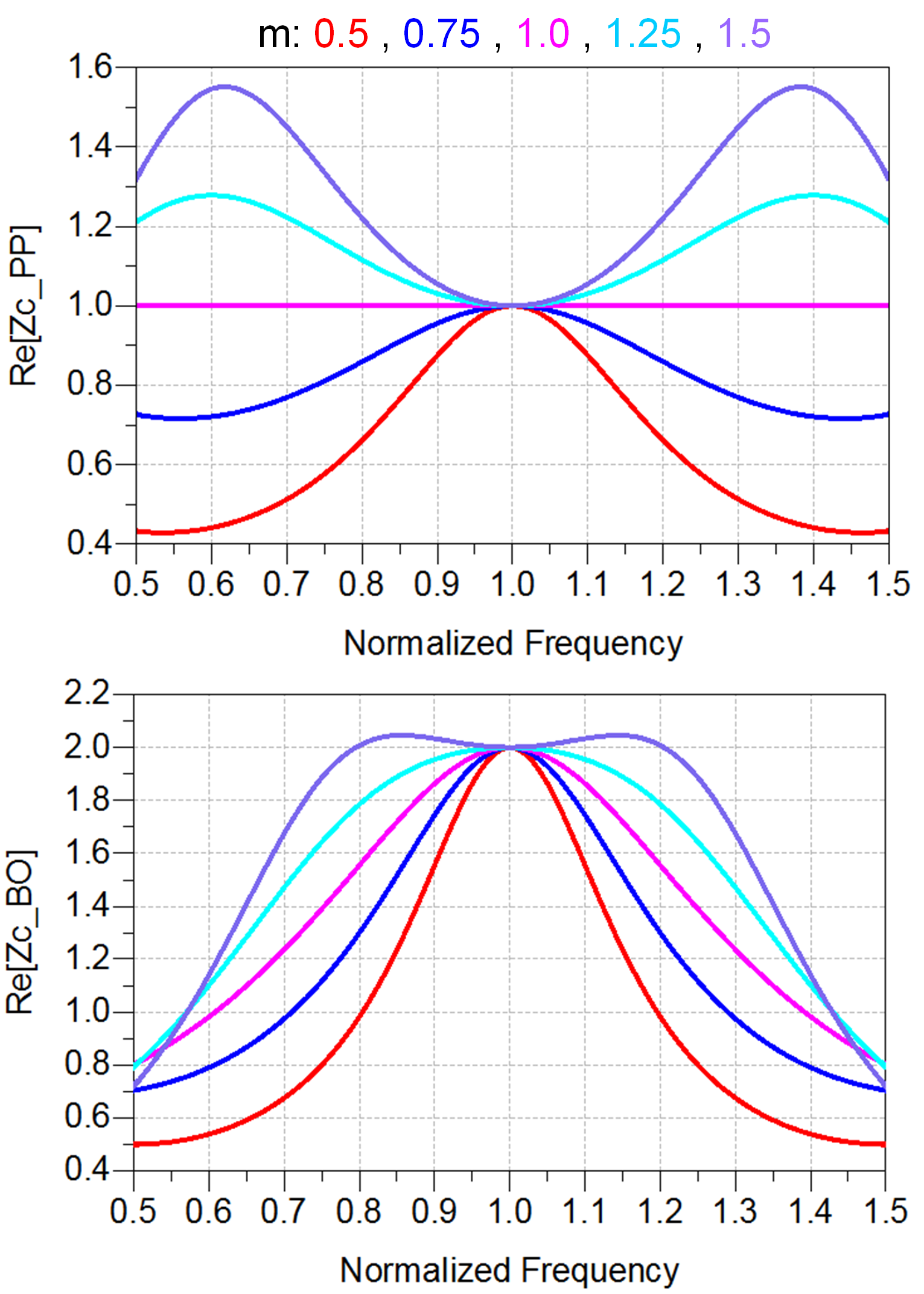}
\caption{Real part of the normalized impedance presented to the carrier amplifier at peak power (top) and 6-dB back-off (bottom) for different values of $m=\sqrt{R_{opt} / 2 R_L}$ (ideally, $m=1$). The bandwidth at back-off can be improved using a proper value of $m$.}
\label{Zc1_v2}
\end{figure}

The transmission line TL$_2$ operates as an impedance matching network that should transform $R_L$ to $R_{opt}/2$ over the bandwidth. It can limit the DPA bandwidth when the impedance transformation ratio, $2R_L/R_{opt}$, is excessive. To investigate the bandwidth limitation effect of TL$_2$, we derive the impedance presented to the carrier amplifier. It can be shown that the impedance $Z_{c}(f)$ can be derived from (\ref{eq_Zc1}) by replacing the parameter $k$ with $k_c$ given by
\begin{equation}
\label{eq_kc}
k_c= k \frac{ \frac{1}{m}+ j \tan (\frac{\pi}{2} \frac{f}{f_0})}{m+j \tan (\frac{\pi}{2} \frac{f}{f_0})},
\end{equation}
where $m=\sqrt{R_{opt} / 2 R_L}$. The real part of the impedance at peak power and back-off is shown in Fig.~\ref{Zc1_v2}. For $m<1$, the bandwidth degrades compared to the ideal case $m=1$, while for $m>1$ higher bandwidth is achieved, but with additional peaks in the real part of the impedance at peak power, leading to variations in the output power and efficiency over the bandwidth. For $m=0.5$, the fractional bandwidth for 20\% reduction in the real part of the impedance at back-off is 18\%, while for $m=1.5$ it is 62\%. This shows the advantage of using transistors with large $R_{opt}$, e.g., GaN devices, in the realization of broadband DPAs. Note that the characteristic impedance of TL$_2$ is given by $Z_2=mR_L$, which can become unfeasible if $m$ is too large.

\begin{figure}[!t]
\centering
\includegraphics[width=2.8in]{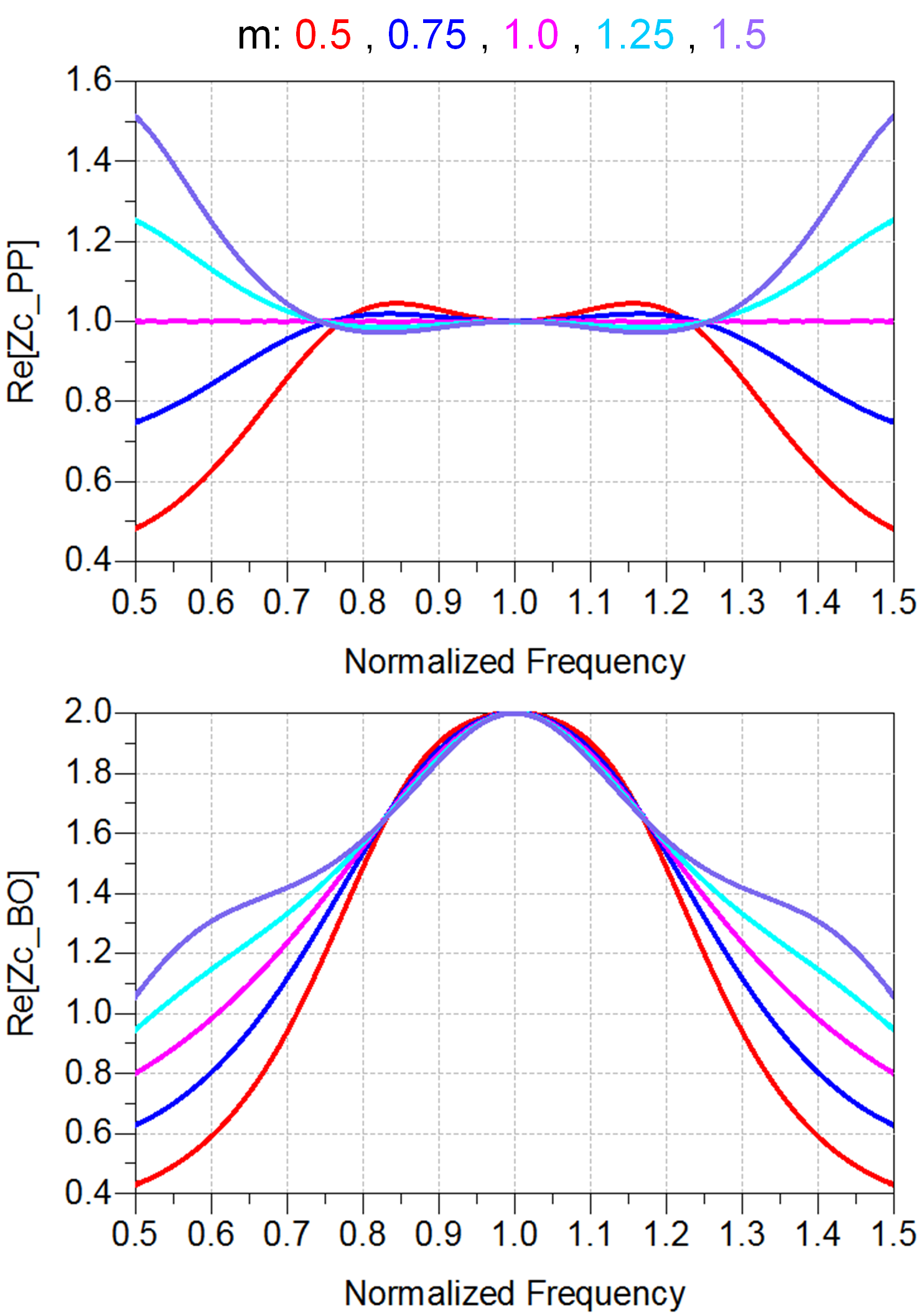}
\caption{Real part of the impedance presented to the carrier amplifier at peak power and 6-dB back-off by replacing TL$_2$ with a two-section impedance transformer composed of two $\lambda/4$ transmission lines.}
\label{Zc1_v3}
\end{figure}

The TL$_2$ transmission line can be replaced by a higher-order matching network to improve its bandwidth. For example, a two-section impedance transformer composed of two $\lambda/4$ transmission lines with the characteristic impedances of $Z_{2a}=\sqrt{m} R_L$ and $Z_{2b}=m\sqrt{m} R_L$ will lower the impedance transformation ratio ($m$ for each transmission line compared to $m^2$ in the previous case) thus extending the bandwidth, as shown in Fig.~\ref{Zc1_v3}. The fractional bandwidth for 20\% reduction in the real part of the impedance at back-off is 38\% for all values of $m$. Also, the real part of the impedance at peak power exhibits much smaller variations over the bandwidth.

It should be noted that this technique is not effective for the transmission line TL$_1$ which should operate as an impedance inverter. To clarify this point, we note that a higher-order matching network is designed for a given set of source and load impedances. If the load resistance is doubled, for example, its frequency response degrades. An impedance inverter, however, is a special impedance transformer in that its input impedance is proportional to the inverse of the load impedance. It is not straightforward to realize a broadband impedance inverter by replacing a $\lambda/4$ transmission line with a higher-order network.

A DPA with asymmetric (i.e., of unequal size) transistors is used to amplify signals with PAPR larger than 6~dB. Assuming the strength of peaking transistor is $N$ times of the carrier transistor, the peak efficiency is achieved at $20 \log (N+1)$-dB back-off. The characteristic impedance of TL$_1$ is the same as in the symmetric DPA, i.e., $Z_1 = R_{opt}$, while for TL$_2$ it is given by
\begin{equation}
\label{eq_z2_N}
Z_2= \sqrt{\frac{1}{N+1} R_L R_{opt}}.
\end{equation}
The load impedance is transformed to $R_{opt}/(N+1)$ at the common node. The impedance presented to the carrier transistor is $R_{opt}$ at peak power and $(N+1)R_{opt}$ at back-off, while the impedance $R_{opt}/N$ is presented to the peaking amplifier at peak power (it is the optimum impedance of the peaking transistor as its size is $N$ times of the carrier transistor). It is emphasized that the impedance transformation ratio of TL$_1$ at back-off is increased to $(N+1) R_L / R_{opt}$, limiting the bandwidth of the asymmetric DPA. Moreover, the impedance transformation ratio for TL$_2$ is $(N+1)R_L/R_{opt}$, which can be larger than that of the symmetric DPA, further limiting the bandwidth.

\subsection{Parasitic Capacitances}

\begin{figure}[t]
\centering
\includegraphics[width=2.4in]{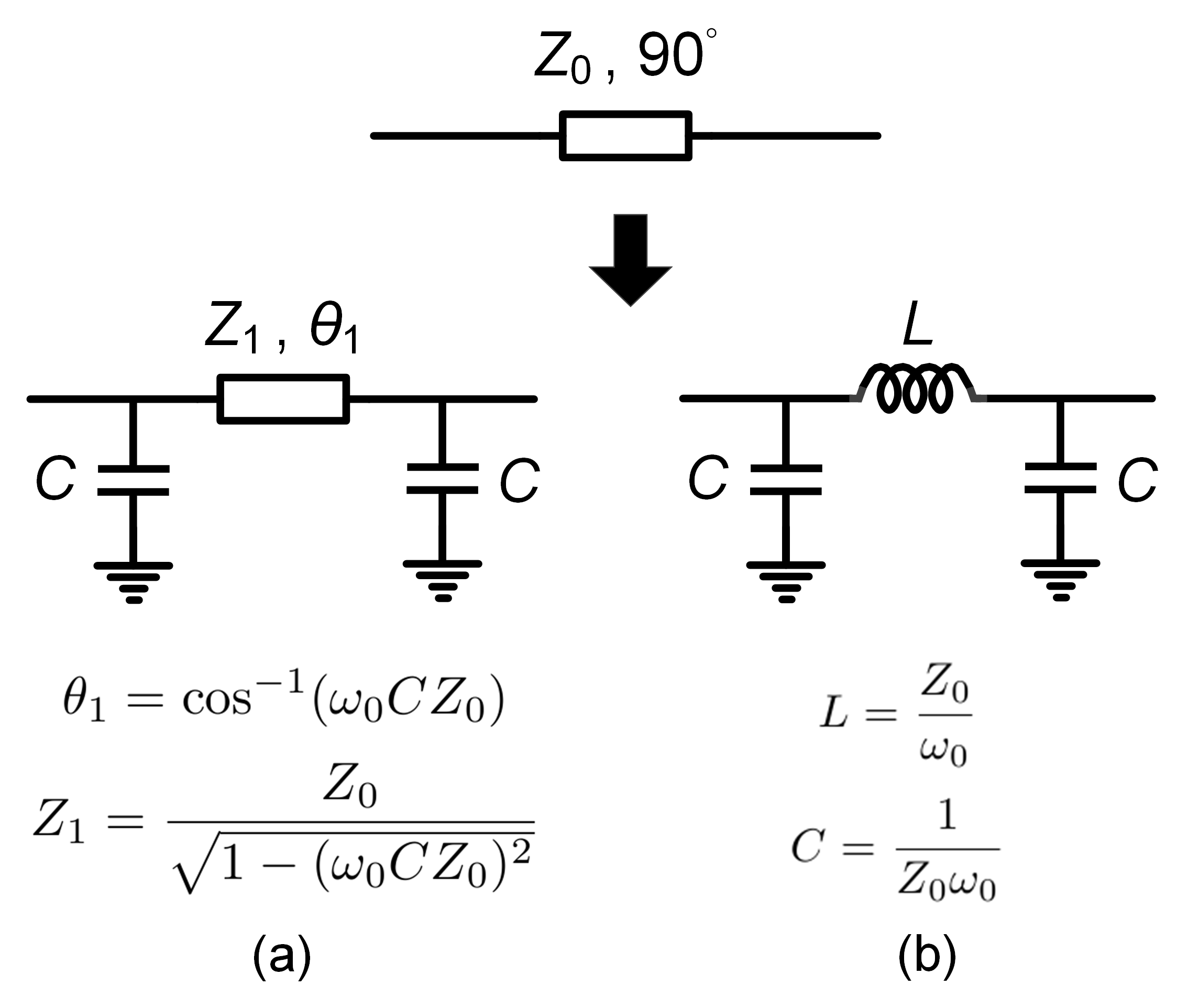}
\caption{Conventional techniques to absorb output parasitic capacitances of transistors into the impedance inverter (a) reduced-length transmission line, (b) lumped-element transmission line.}
\label{parasitic-absorption}
\end{figure}

Parasitic capacitances of the transistors can limit the DPA bandwidth. The drain-source capacitance $C_{ds}$ can affect the impedance presented to the amplifiers thus limiting the bandwidth of the load modulation network. A solution is to absorb the transistors' parasitic capacitances into the impedance inverter network using the reduced-length or lumped-element equivalent circuit of a transmission line, as shown in Fig.~\ref{parasitic-absorption}. Unfortunately, both equivalent circuits are only valid at a single frequency and their frequency response deviates from the original circuit. Moreover, there is a limit on the parasitic capacitance that can be absorbed by the impedance inverter circuit. In Fig.~\ref{parasitic-absorption}(a), a large capacitance leads to a large characteristic impedance and unrealistic transmission line, while in Fig.~\ref{parasitic-absorption}(b) $C_{ds}$ should be smaller than $C=1/Z_0 \omega_0$. We investigate the impedance presented to the carrier amplifier in these two cases and compare the results with a transmission line based impedance inverter.

\begin{figure}[!t]
\centering
\includegraphics[width=2.8in]{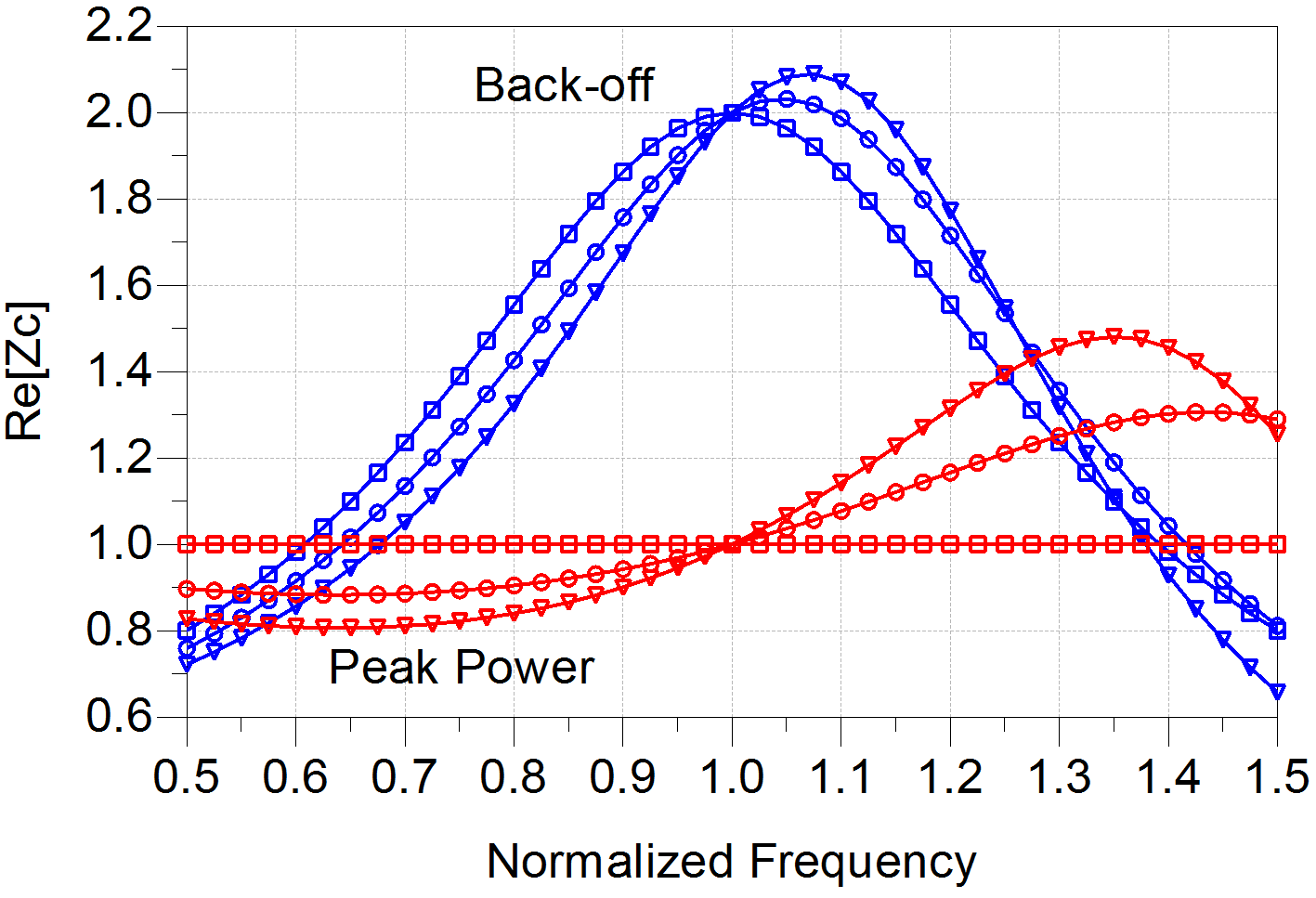}
\caption{The real part of the impedance presented to the carrier amplifier using a reduced-length transmission line for parasitic absorption. Normalized parasitic capacitance $\omega_0 C R_{opt}$ is 0 (square), 0.25 (circle), 0.5 (triangle).}
\label{Zc2}
\end{figure}

The real part of the impedance in the former case is plotted in Fig.~\ref{Zc2} for different parasitic capacitances. The circuit bandwidth degrades as the capacitance \textit{C} increases. Moreover, this technique requires a transmission line with higher characteristic impedance, i.e., narrower width, which may not be feasible due to process limitations. To effectively use this circuit in an asymmetric DPA (i.e., with stronger peaking transistor), an extra capacitance should be added to the output of the carrier transistor to equalize the two parasitic capacitances, thus further limiting the bandwidth.

\begin{figure}[!t]
\centering
\includegraphics[width=2.8in]{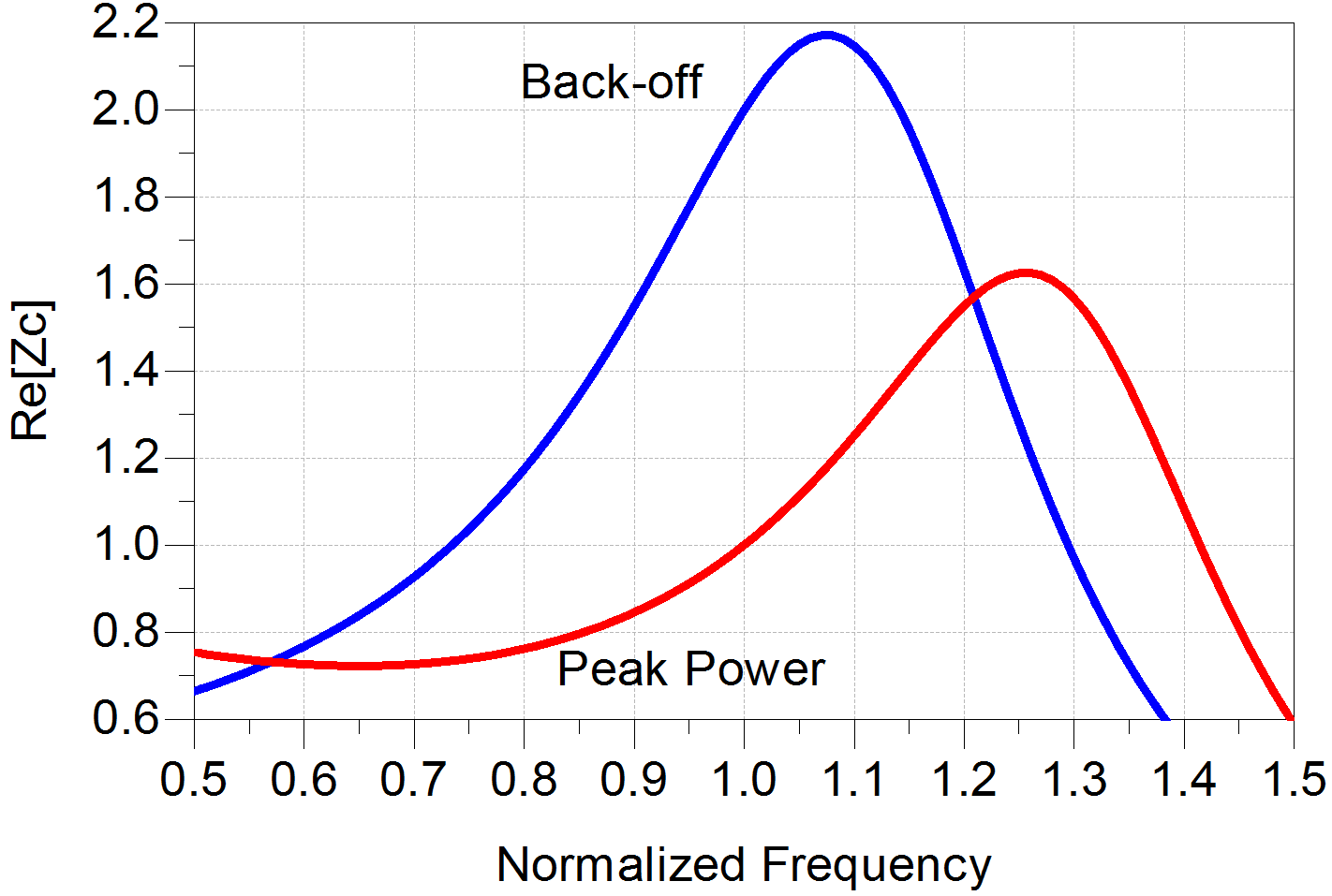}
\caption{Real part of the impedance presented to the carrier amplifier using a lumped-element transmission line. The DPA bandwidth is limited now by both the peak power and back-off impedances.}
\label{Zc3}
\end{figure}

In the second case of Fig.~\ref{parasitic-absorption}(b), the real part of the $Z_c$ impedance is depicted in Fig.~\ref{Zc3}, indicating bandwidth at peak power and back-off is reduced compared to the results in Fig.~\ref{Zc1}.
Other circuit techniques, including offset lines, resonant circuits, and compensation networks, have been proposed to cancel the $C_{ds}$ effects, but these circuits are usually narrowband \cite{grebennikov12}, \cite{camarchia15}, \cite{cidronali16}.

The gate-source capacitance $C_{gs}$ also limits the bandwidth of the input power splitter and phase alignment network. The impedance matching network used to match the input impedance of the transistors to the power splitter can limit the bandwidth for large $C_{gs}$. The $C_{gs}$ nonlinearity, especially for the class-C biased peaking transistor, leads to a dependency of the input impedance (and, as a result, of the input power division ratio) on the input power. This will degrade the DPA linearity \cite{nunes14}.

The gate-drain capacitance $C_{gd}$ presents a nonlinear impedance at the input of the transistors through the Miller's effect. The impedance nonlinearity is more severe for the carrier amplifier whose load impedance varies by a factor of 2, i.e., from $R_{opt}$ at the peak power to $2R_{opt}$ at the 6-dB back-off. Therefore, $C_{gd}$ can limit the bandwidth and degrade linearity of the DPA \cite{piazzon14}.

\subsection{Input Network}

The input power splitter and phase shift network can constitute another source of bandwidth limitation. In a separate implementation of the power splitter, e.g., a Wilkinson power splitter, and phase shift network, e.g., a $\lambda/4$ transmission line, bandwidth is mainly limited by a narrowband phase response of the $\lambda/4$ transmission line. A wider bandwidth can be achieved by merging the two functions in a quadrature coupler, e.g., a Lange coupler or branch line coupler. Further bandwidth enhancement can be achieved by using multi-section Wilkinson power divider or branch line coupler. However, this implementation increases the circuit size and may become infeasible for IC implementations at low frequencies.

\section{DPA Bandwidth Extension Techniques}
\label{sec:BW_extension}

In this section, we provide a review of bandwidth extension techniques for the DPA. In Table \ref{table_DPAs}, a performance summary is presented for broadband DPAs using various bandwidth extension techniques.

\begin{table*}[!t]\footnotesize
\renewcommand{\arraystretch}{1.2}
\setlength{\tabcolsep}{10pt}
\caption{Performance of Broadband DPAs Using Various Bandwidth Extension Techniques.}
\label{table_DPAs}
\centering
\begin{minipage}{7.0in}
\centering
\begin{tabular}{ccccccc}
\hline
 Ref. & Bandwidth& Peak P$_{out}$ & $\eta_{PP}$ & $\eta_{BO}$ & BO &Transistor \\
\hline\hline

\multicolumn{7}{l}{\textit{Modified Impedance of Transmission Lines}} \\
\cite{bathich11} & 1.7--2.6 GHz (42\%) & 42--45 dBm & 55--60\% & 41--55\% & 6 dB &GaN \\
\cite{wu12} & 0.7--1.0 GHz (35\%) & 49--51 dBm & 60--75\% & 52--68\%& 6 dB &GaN \\
\cite{gustafsson13} & 1.5--2.4 GHz (47\%) & 42 dBm & 52--66\% & 49--62\% & 6 dB &GaN \\
\cite{darraji17} & 0.55--1.1 GHz (67\%) & 42--43.5 dBm & 56--72\% & 40--52\% & 6 dB & GaN \\
\hline
\multicolumn{7}{l}{\textit{Two-Section Peaking Network}} \\
\cite{grebennikov12-2} & 2.1--2.7 GHz (25\%) & 42.7--43.8 dBm & 68--74\% & 52--58\% & 5--6 dB &GaN \\
 \cite{giofre14-2} & 1.0--2.6 GHz (83\%) & 40--42 dBm & 45--83\% & 35--58\% & 6 dB &GaN \\
\cite{barakat17} & 2.1--2.66 GHz (24\%) & 43 dBm & 57--84\% & 39--67\% & 6 dB & GaN \\
\hline
\multicolumn{7}{l}{\textit{Networks Based on Branch-Line Couplers}} \\
 \cite{giofre13} & 1.95--2.25 GHz (14\%) & 41.5--42 dBm & 65--65\% & 48--50\% & 6 dB &GaN \\
 \cite{piazzon13} & 1.67--2.41 GHz (36\%) & 39--41 dBm & 53--72\%& 43--59\% & 6 dB &GaN \\
 \hline
 \multicolumn{7}{l}{\textit{Peaking Network With Shunt Short-Circuited Stub}} \\
 \cite{chen16} & 1.5--2.5 GHz (52\%) & 42--44.5 dBm & 55--75\%& 42--53\% & 6 dB &GaN \\
 \cite{chen16-2} & 2.0--2.6 GHz (26\%) & 44--45 dBm & 53-76\% & 40--47\% & 8 dB &GaN \\
\hline
\multicolumn{7}{l}{\textit{Frequency Response Optimization}} \\
\cite{sun12} & 2.2--2.96 GHz (30\%) &40--41.5 dBm & 52--68\% & 40--47\% & 5--6 dB &GaN \\
\cite{bathich12} & 1.7--2.25 GHz (28\%) & 48.2--49.6 dBm & 65--77\% & 53--65\% & 6 dB &GaN \\
\cite{chen17} & 1.5--2.4 GHz (47\%) & 43--44 dBm & 57--75\% & 45--54\% & 7 dB &GaN \\
\hline
\multicolumn{7}{l}{\textit{Parasitic Compensation}} \\
\cite{rubio12} & 3.0--3.6 GHz (18\%) & 43--44 dBm & 55--66\% & 38--56\%& 6 dB & GaN \\
\cite{rubio18} & 1.5--3.8 GHz (87\%) & 42.3--43.4 dBm & 42--63\% & 33--55\%& 6 dB & GaN \\
\cite{cidronali16} & 0.65--0.95 GHz (37.5\%) & 54 dBm & 51--72\%& 50--60\% & 4 dB & LDMOS \\
\cite{xia16-2} & 1.7--2.8 GHz (49\%) & 44--44.5 dBm & 57--71\%& 50--55\% & 6 dB & GaN \\
\hline
\multicolumn{7}{l}{\textit{Post-Matching DPA}} \\
\cite{pang15} & 1.7--2.6 GHz (43\%)& 44--46 dBm& 57--66\% & 47--57\% & 6 dB &GaN \\
\cite{zhou17} & 1.8--2.7 GHz (40\%) & 41 dBm & 57--73\% & 47--54\% & 6 dB &GaN \\
\cite{kang18} & 0.9--1.8 GHz (70.7\%) & 49.7--51.4 dBm & 54--73\% & 42--58\% & 6 dB & GaN \\
\cite{chen16-3} & 1.65--2.75 GHz (50\%) & 44--46 dBm & 60--77\% & 52--66\% & 6 dB &GaN \\
\cite{shi18} & 1.6--2.7 GHz (53\%) & 43.8--45.2 dBm & 56--75\% & 46--63\% & 6 dB & GaN \\
\hline
\multicolumn{7}{l}{\textit{Dual-Input DPA}} \\
 \cite{andersson13} & 1--3 GHz (100\%) & 43--45 dBm & 45--68\% & 48--68\% & 6 dB &GaN \\
 \hline
 \multicolumn{7}{l}{\textit{Transformer-Less Load Modulation PA}} \\
\cite{akbarpour12} & 2.0--2.45 GHz (20\%)& 40--42 dBm & 56--67\% & 40-45\% & 6 dB &GaN \\
\cite{watanabe15} & 1.6--2.0 GHz (22\%) & 31--34 dBm & 47--64\% & 20--50\% & 6 dB &GaN \\

\hline
\end{tabular}
\end{minipage}
\end{table*}

\subsection{DPA with Modified Load Modulation Network}

\subsubsection{Modified Impedance of Transmission Lines}

As stated earlier, in the conventional DPA architecture, assuming symmetrical transistors, the impedance transformation ratio of the impedance inverter is 1 at peak power and 4 at 6-dB back-off, thus limiting the DPA bandwidth at back-off. A number of DPA architectures with modified characteristic impedance of transmission lines were proposed to mitigate this issue.

In \cite{bathich11}, the common load impedance was increased from $R_{opt}/2$ to a higher value, e.g., $\sqrt{2} R_{opt}/2$. This reduces the impedance transformation ratio at 6-dB back-off to $2\sqrt{2} \cong 2.8$. Therefore, the drain efficiency bandwidth extends as compared to that in the conventional DPA. The characteristic impedances of the transmission lines are chosen as $Z_1= \sqrt[4]{2} R_{opt}$ and $Z_2= \sqrt{R_L R_{opt}}/ \sqrt[4]{2}$. The load impedance presented to the peaking amplifier at saturation is $\sqrt{2} R_{opt}$, which is higher than its optimum value. This leads to the degradation of output power and efficiency at saturation. A GaN DPA designed using this technique achieves 41--55\% drain efficiency at 6-dB back-off in 1.7--2.6 GHz (42\%).

\begin{figure}[!t]
\centering
\includegraphics[width=2.5in]{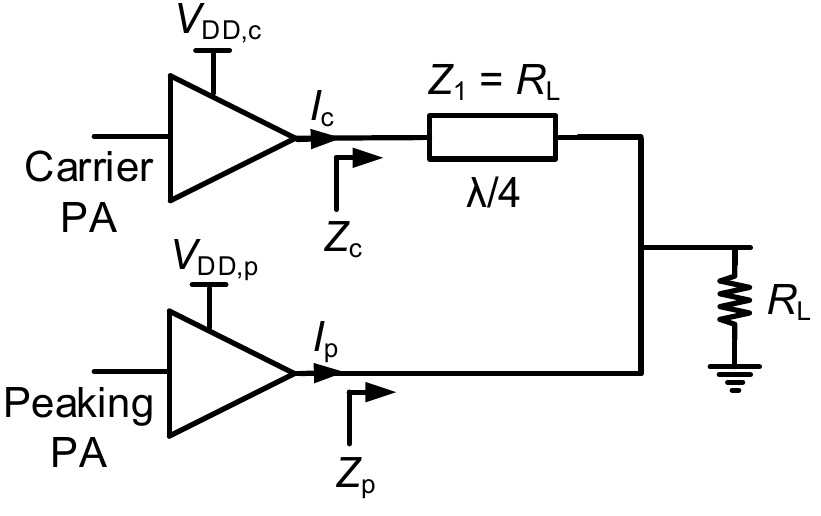}
\caption{Modified DPA architecture with improved back-off bandwidth proposed in \cite{wu12}, \cite{gustafsson13}.}
\label{modified-DPA}
\end{figure}

A modified DPA architecture was proposed in \cite{wu12}, \cite{gustafsson13}, where the characteristic impedance of the impedance inverter is chosen identical to the load resistance, $Z_1=R_L$, as shown in Fig.~\ref{modified-DPA}. A frequency-independent impedance is thus presented to the carrier amplifier at back-off, thus improving the DPA bandwidth. The impedances $Z_c$ and $Z_p$ are manipulated to achieve broadband performance at both the peak power and back-off by proper control of the relative phase and amplitude of the carrier and peaking amplifier currents, $I_c$ and $I_p$. Asymmetric drain supply voltages are used for the carrier and peaking amplifiers, and it is shown that the power back-off level can be reconfigured by the ratio of two supply voltages. Using this technique, a GaN DPA was presented in \cite{wu12} with 52--68\% efficiency at 6-dB back-off, in 0.7--1.0 GHz (35\%). In \cite{gustafsson13}, a broadband GaN DPA with two RF inputs was reported, achieving 49--62\% efficiency at 6-dB back-off in 1.5--2.4 GHz (47\%). The need for unequal drain bias voltages is a drawback of this architecture, as one of the transistors should operate at a supply voltage lower than the maximum level allowed by the process. However, the advantage is being able to reconfigure the back-off level to accommodate modulated signals with different PAPR.

\begin{figure}[!t]
\centering
\includegraphics[width=3.0in]{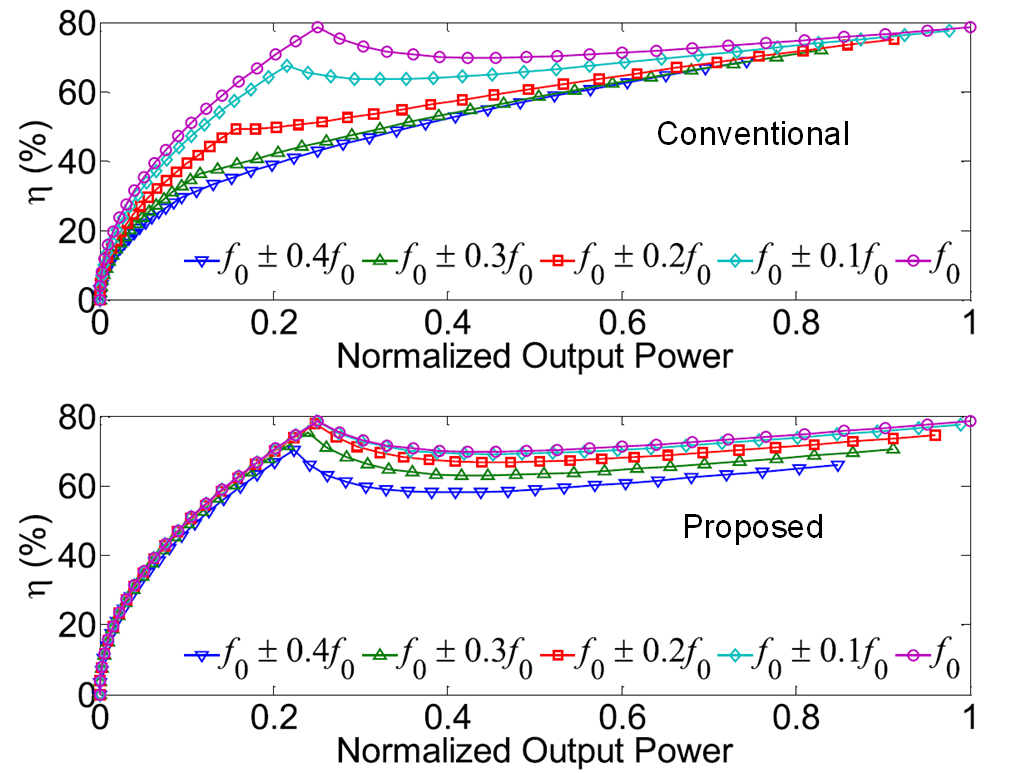}
\caption{Theoretical drain efficiency of the DPA with maximally flat frequency response compared to the conventional DPA \cite{darraji17}.}
\label{max-flat-DPA}
\end{figure}

Another broadband DPA design approach was presented in \cite{darraji17}, where the characteristic impedances of the two transmission lines are determined for impedance transformation with maximally flat frequency response. The resulting impedances are derived as
\begin{equation}
\label{eq_z1_opt}
Z_1=\sqrt{\frac{1+1/\sqrt{\alpha}}{2\alpha}} R_{opt}
\end{equation}
\begin{equation}
\label{eq_z2_opt}
Z_2= \sqrt{\frac{1+1/\sqrt{\alpha}}{2}} R_{opt},
\end{equation}
where $\alpha$ is the back-off level (0.5 for the 6-dB back-off). The phase and magnitude of the current profiles are set based on the input power and frequency. This architecture requires asymmetric supply voltages for the carrier and peaking amplifier, related as
\begin{equation}
\label{eq_VDD_ratio}
V_{DD,p} = \sqrt{\frac{1+1/\sqrt{\alpha}}{2\alpha}} V_{DD,c}.
\end{equation}
The theoretical efficiency of this DPA architecture is compared with the conventional DPA in Fig.~\ref{max-flat-DPA}, where it is noticed that the DPA with maximally flat frequency response provides notable efficiency enhancement at back-off. A GaN DPA with two RF inputs is implemented based on this architecture. It exhibits drain efficiency of 40--52\% at 6-dB back-off in 0.55--1.10 GHz (67\%). The main advantage there is the broadband operation achieved through using optimum carrier and peaking impedances, while the need for asymmetric supply voltages limits the power capability of the carrier transistor.

\subsubsection{Two-Section Peaking Network}

\begin{figure}[!t]
\centering
\includegraphics[width=3.0in]{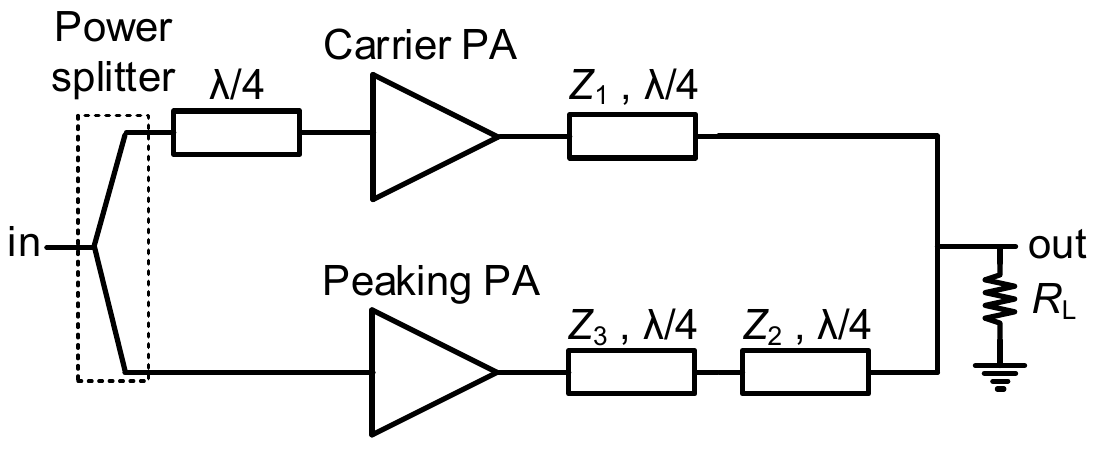}
\caption{The DPA architecture with two-section peaking network for bandwidth extension at back-off.}
\label{two-section-peak}
\end{figure}

Yet another modified DPA architecture with two $\lambda/4$ transmission lines in the peaking network is shown in Fig.~\ref{two-section-peak}. The output impedance transformation network is eliminated, while characteristic impedances of the transmission lines are established to achieve a broadband response. This architecture has been extensively investigated and different design criteria have been derived for its optimal operation \cite{grebennikov12-2}-\cite{barakat17}.

In \cite{grebennikov12-2}, the characteristic impedances of the transmission lines, assuming $R_L$ and $R_{opt}$ of 50 $\Omega$, are chosen as $Z_1=Z_2=50\sqrt{2}\cong 70.7~\Omega$ and $Z_3=50~\Omega$. In the proposed architecture, the impedance inverter TL$_1$ performs a 100-to-50 $\Omega$ transformation at peak power and a 50-to-100 $\Omega$ transformation at 6-dB back-off. Therefore, the impedance transformation ratio is 2 in both cases, leading to an extended bandwidth at back-off but reduced bandwidth at peak power. A broadband DPA with 40--45\% drain efficiency at 6.5~dB back-off and 2.1--2.7 GHz (25\%) bandwidth was designed using this architecture based on GaN class-E amplifiers.

In \cite{barakat17}, a generalized design approach was presented for this DPA architecture. Assuming transistors with the same optimum load resistance, it is shown that to achieve broadband efficiency at back-off and peak power, the characteristic impedances of the transmission lines should be chosen as

\begin{equation}
\label{eq_z1}
Z_1 = \frac{R_{opt}}{m}
\end{equation}
\begin{equation}
\label{eq_z2}
Z_2 = \frac{R_{opt}}{m \sqrt{m}}
\end{equation}
\begin{equation}
\label{eq_z3}
Z_3 = \frac{R_{opt}}{\sqrt{m}}
\end{equation}
where $m = \sqrt{R_{opt}/2R_L}$. The impedance transformation ratio for TL$_1$ is derived as $m^2$ at peak power and $4m^2$ at 6-dB back-off, while for TL$_{2,3}$ it is given by $m$ at peak power. It is noticed that for $m=1$, this architecture provides the same results as the conventional DPA, while for $m=0.5$ the largest bandwidth is obtained at back-off. The DPAs reported in \cite{wu14}, \cite{giofre14}, \cite{giofre14-2} are special cases of this architecture with $m$ of 0.6, 0.6, and 0.4, respectively. A GaN DPA presented in \cite{barakat17} was based on a class-E PA and the output combiner with $m=0.5$, achieving drain efficiency of 39--67\% at 6-dB back-off in 2.1--2.66 GHz (24\%). In \cite{giofre14-2}, \cite{giofre14-3} a GaN DPA is designed using this architecture with drain efficiency of 35--58\% at 6-dB back-off in 1.0--2.6 GHz (83\%). A broad bandwidth is achieved, but with large drain efficiency variations over the bandwidth. This indicates that the load modulation network is not capable of providing a constant optimum load resistance over the bandwidth for the transistors. This architecture was used in \cite{he15} to realize an LDMOS broadband push-pull DPA.
The DPA provides 700~W output power and over 40\% average drain efficiency for a signal with 7.5 dB PAPR in the bandwidth of 522--762 MHz (37\%). This architecture has a straightforward design procedure and can provide wide bandwidth at back-off. Its drawbacks are large variations of the drain efficiency within the bandwidth and the extra quarter-wavelength transmission line.

More complicated load modulation networks based on branch-line couplers were proposed in \cite{giofre13} and \cite{piazzon13}. However, as can be noticed in Table \ref{table_DPAs}, the reported bandwidths are lower than those achieved using the two-section peaking network with a simpler structure.

\subsubsection{Peaking Network With Shunt Short-Circuited Stub}

\begin{figure}[!t]
\centering
\includegraphics[width=2.0in]{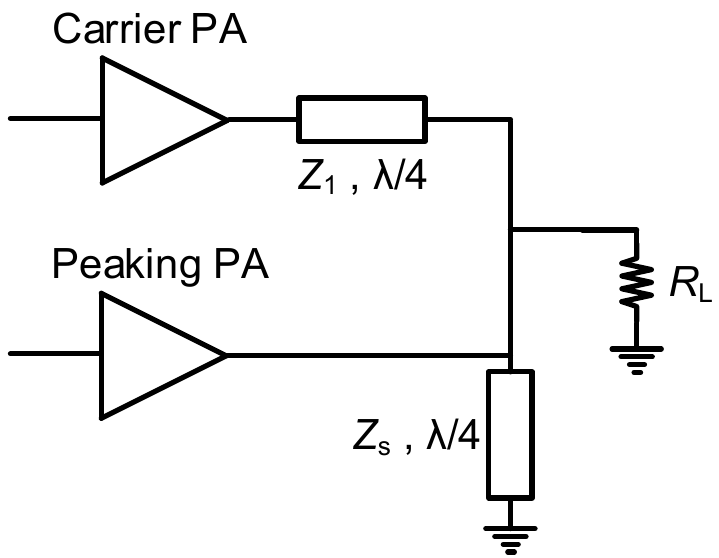}
\caption{The DPA bandwidth extension using short-circuited stub \cite{chen16}.}
\label{sc-stub}
\end{figure}

\begin{figure}[!t]
\centering
\includegraphics[width=2.8in]{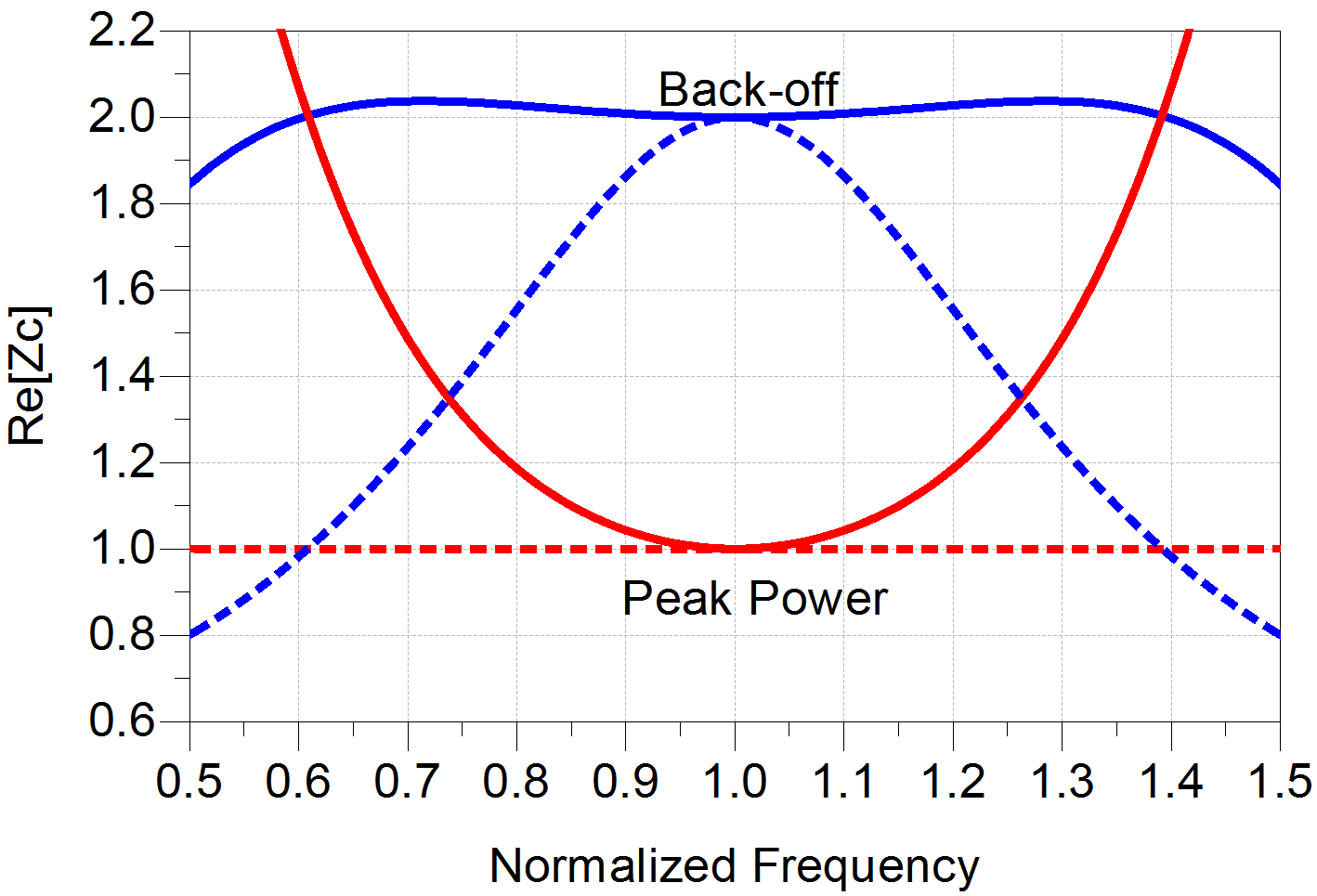}
\caption{The impedance presented to the carrier amplifier of the DPA with short-circuited stub at peak power and 6-dB back-off (solid lines) compared to the conventional DPA (dashed lines).}
\label{sc-stub-impedance}
\end{figure}

In \cite{chen16}, a $\lambda/4$ shunt short-circuited stub was added at the output of peaking amplifier to improve the bandwidth of the load modulation network, as shown in Fig.~\ref{sc-stub}. This stub modifies the impedance presented to the impedance-inverting transmission line in order to compensate for the drop in impedance seen by the carrier amplifier. The characteristic impedances of the impedance inverter and the short-circuited stub are chosen as $Z_1=50 \sqrt{2}~\Omega$ and $Z_s=120~\Omega$. Simulation results for the carrier impedance $Z_c$ at the back-off and peak power for the proposed and conventional DPA architectures are compared in Fig.~\ref{sc-stub-impedance}. It is noticed that the real part of the carrier impedance exhibits significant bandwidth improvement in back-off. The larger impedance at band edges can degrade the DPA efficiency at peak power. A GaN DPA designed using this technique provides 42--53\% drain efficiency at 6-dB back-off, in 1.5--2.5 GHz (52\%). Another GaN DPA designed based on this technique, with two peaking amplifiers and a carrier amplifier, achieves 40--47\% drain efficiency at 8-dB back-off in 2.0--2.6 GHz (26\%) \cite{chen16-2}. The main advantage of this architecture is that the broadband operation can be achieved by using only an extra transmission line. However, the characteristic impedance of this transmission line may become too large, thus impractical.


A similar DPA architecture was proposed in \cite{abadi14}, where a parallel LC network is used instead of the short-circuit stub. The parallel network resonates at the center of the frequency band, while it improves bandwidth of the carrier impedance at back-off. The DPA exhibits 48--56\% drain efficiency at the 6-dB back-off in 700--950 MHz (30\%).

\subsection{Frequency Response Optimization}

\begin{figure}[!t]
\centering
\includegraphics[width=2.8in]{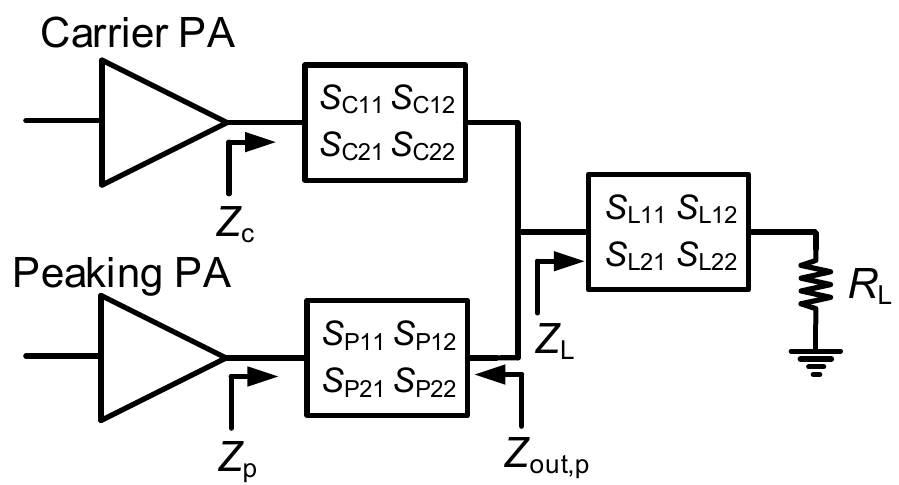}
\caption{The DPA model based on two-port networks for real frequency design technique \cite{sun12}.}
\label{SRF-DPA}
\end{figure}

A broadband design approach for the DPA is based on optimization of the overall frequency response \cite{sun12}, \cite{bathich12}, \cite{chen17}.
In \cite{sun12}, a ``real frequency" technique is used for synthesis of matching networks. The DPA output network is represented by three two-port networks, shown in Fig.~\ref{SRF-DPA}. Their scattering parameters are determined based on conditions that should be satisfied by the impedances at both saturation and back-off. If a high back-off efficiency is the most important design target, the optimum load impedance, $Z_{c,opt,BO}$, should be presented to the transistors at back-off, while the impedances at saturation can only reach a sub-optimum value. The load impedance, $Z_L$, should be optimized to provide a broadband response to the carrier amplifier at back-off. The output impedance of the peaking amplifier network, $Z_{out,p}$, should be close to open-circuit to avoid power leakage from the carrier amplifier at back-off. Using this technique, a broadband GaN DPA is designed in which a drain efficiency of 40--47\% at 5--6~dB back-off is achieved in 2.2--2.96 GHz (30\%).

In \cite{bathich12}, the frequency-dependent back-off efficiency degradation is minimized by a proper design of the carrier amplifier output matching network to compensate for the effect of frequency-sensitive impedance of the inverter. The impedance presented to the carrier amplifier at back-off exhibits a broadband real part. A GaN DPA designed based on this technique provides 53--65\% drain efficiency at the 6-dB back-off in 1.7--2.25 GHz (28\%).

In \cite{chen17}, a multi-objective Bayesian optimization is used to design a broadband GaN DPA. It is shown that this optimization strategy outperforms a built-in optimizer of commercial electronic design tool. The DPA achieves 20-W output power, 45--54\% average drain efficiency at 7~dB back-off for a 20-MHz single carrier LTE signal in the bandwidth of 1.5--2.4 GHz (47\%).

\subsection{Parasitic Compensation}

\begin{figure}[!t]
\centering
\includegraphics[width=1.7in]{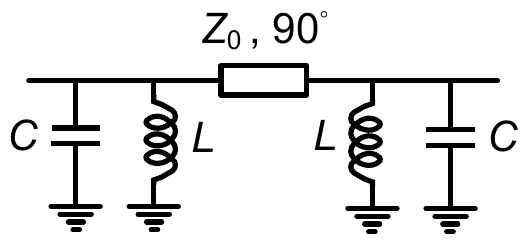}
\caption{Parasitic capacitance cancellation using parallel inductors.}
\label{parasitic-cancellation}
\end{figure}

\begin{figure*}[!t]
\centering
\includegraphics[width=5.0in]{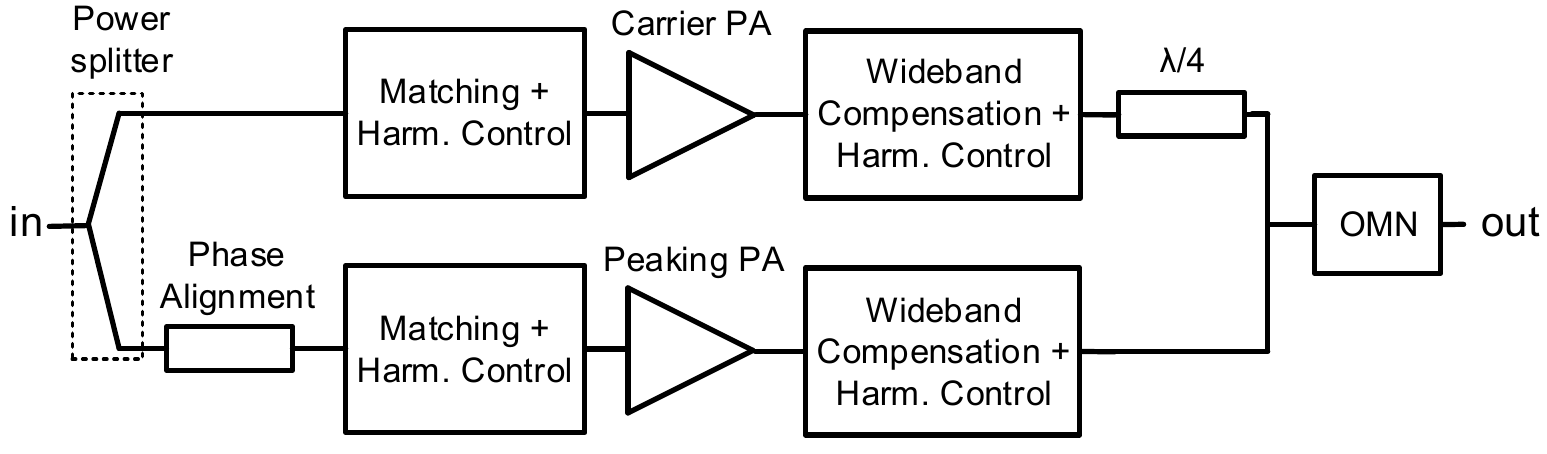}
\caption{The DPA architecture with parasitic compensation and second-harmonic control \cite{rubio12}.}
\label{compensation}
\end{figure*}

The bandwidth of the DPA can be limited by parasitic capacitances of the transistors. Furthermore, output parasitic components alter impedances presented to the intrinsic drain nodes of transistors, leading to variations of the output power and efficiency over the bandwidth. Several parasitic compensation techniques have been presented for the DPA \cite{qureshi10}--\cite{xia16-2}. A basic solution, discussed before, is to absorb the output parasitic capacitances into the impedance inverter network using the reduced-length or lumped-element equivalent circuit of a transmission line. Another technique is to use parallel inductors to resonate with the parasitic capacitances (see Fig.~\ref{parasitic-cancellation}). However, the cancellation is achieved only at a single frequency and loss of the inductors can degrade the DPA efficiency.

In \cite{rubio12}, wideband reactive networks are cascaded with the output of the carrier and peaking amplifiers to compensate for their output parasitic components, shown in Fig.~\ref{compensation}. The compensation network is designed such that the overall two-port network (i.e., the cascaded networks) provides the scattering parameters of $S_{11}=S_{22}=0$ and $S_{21}=S_{12}=\pm 1$. Nevertheless, achieving these conditions over a wide bandwidth, especially the phase response, is not trivial. The implemented GaN DPA achieves the drain efficiency of 38--56\% at the 6-dB back-off in 3.0--3.6 GHz (18\%). Another GaN DPA with parasitic compensation is presented in \cite{rubio18}, which achieves 33--55\% drain efficiency at 6-dB back-off in 1.5--3.8 GHz (87\%). It is noted that the efficiency exhibits large variations in the reported bandwidth.

\begin{figure}[!t]
\centering
\includegraphics[width=2.2in]{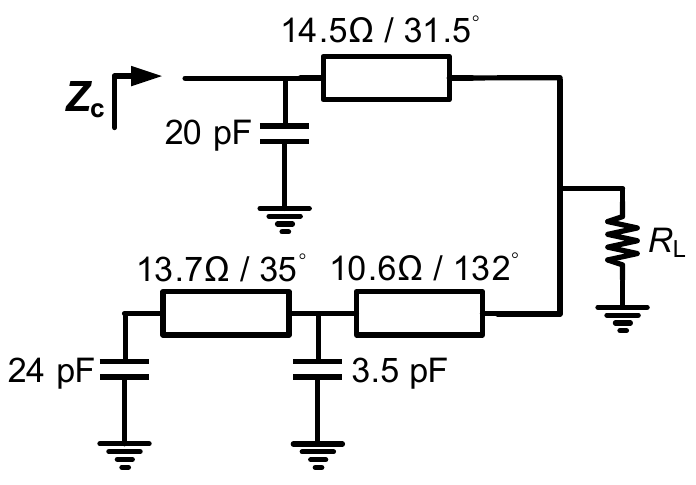}
\caption{Parasitic-compensated load modulation network proposed in \cite{cidronali16}.}
\label{parasitic-compensated-inverter}
\end{figure}

A design technique for a broadband impedance inverter in the presence of large parasitic capacitances was proposed in \cite{cidronali16}. Shown in Fig.~\ref{parasitic-compensated-inverter}, the DPA architecture is based on the two-section peaking network previously discussed. In the conventional parasitic absorption technique, the parasitic capacitances of the carrier and peaking transistors are absorbed into the transmission lines TL$_1$ and TL$_2$ by shortening their length and increasing their characteristic impedance. It is proposed there to replace the transmission line TL$_3$ with an equivalent circuit of two shunt inductors and a transmission line with longer length. This topological transformation is performed such that the extra inductors can compensate for the extra capacitance introduced by shortening the length of transmission lines TL$_1$ and TL$_2$. It is shown that this cancellation can be obtained over a wide bandwidth. A DPA designed based on this technique, and implemented using silicon LDMOS transistors, features average drain efficiency of 37--47\% and average output power of 49~dBm in 650--950 MHz (37.5\%), for a 20-MHz LTE signal with 7.5~dB PAPR.

\subsection{Post-Matching DPA}

\begin{figure}[!t]
\centering
\includegraphics[width=3.3in]{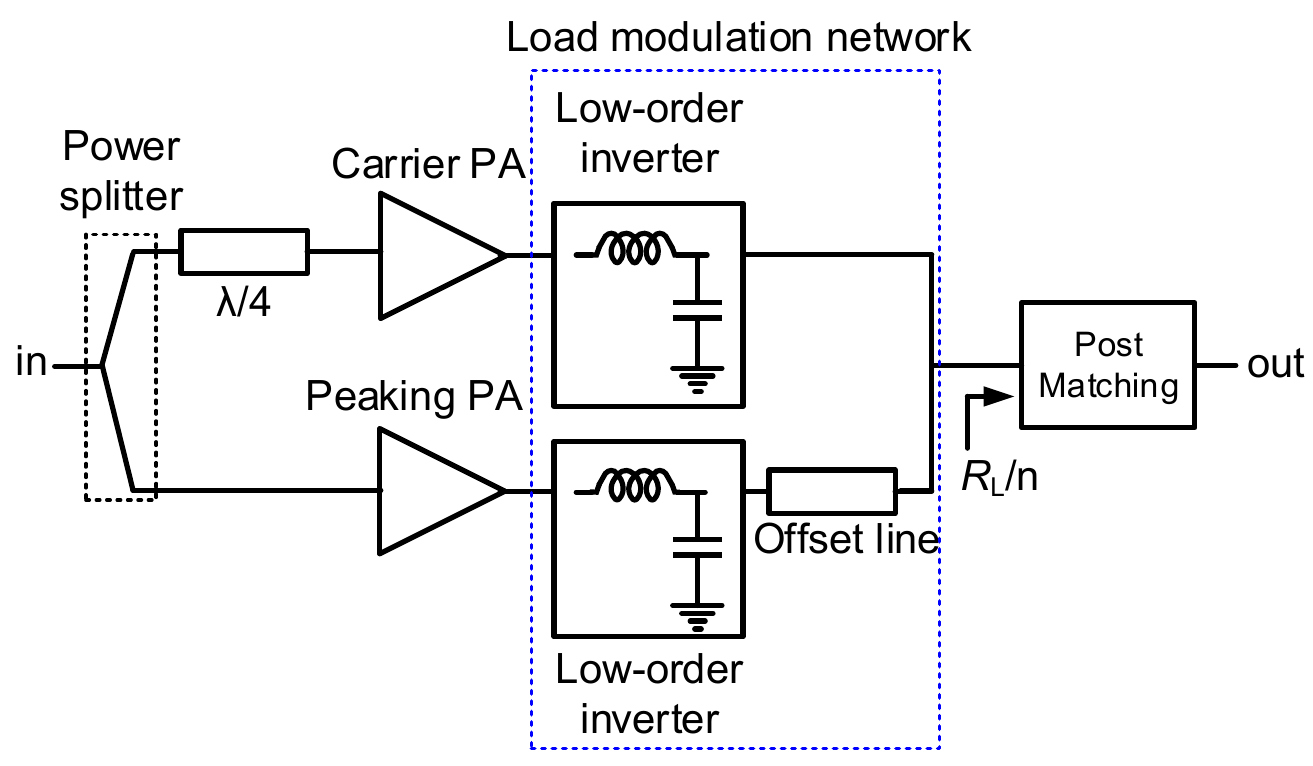}
\caption{The post-matching DPA proposed in \cite{pang15}.}
\label{post-matching-DPA}
\end{figure}

In the post-matching DPA architecture proposed in \cite{pang15}, the impedance matching networks of the carrier and peaking amplifiers are realized by simple low-pass networks in order to extend the bandwidth. Furthermore, a broadband impedance matching network is used at the output of DPA to transform the load resistance into the optimum resistance for broadband operation (Fig.~\ref{post-matching-DPA}). This is different from the conventional impedance matching network at the DPA output which transforms the 50-$\Omega$ load resistance to a fixed resistance (e.g., $2R_{opt}$). This post-matching network provides an appropriate frequency-dependent impedance for the low-order impedance inverters. The implemented GaN DPA achieves drain efficiency of 47--57\% at 6-dB back-off in 1.7--2.6 GHz (43\%). Modulation measurements using a 20-MHz LTE signal with 10.5~dB PAPR indicate higher than 40\% average drain efficiency. In \cite{zhou17}, second-harmonic short-circuit networks are included into the post-matching DPA to improve efficiency. The implemented GaN DPA provides drain efficiency of 47--54\% at 6-dB back-off in 1.8--2.7 GHz (40\%).

A modified version of this architecture was proposed in \cite{kang18}, where a multiple-resonance circuit is used for the peaking amplifier. The resonant network is designed to provide an optimum load resistance to the peaking transistor at the band center, while providing an optimum susceptance $B_{opt}(\omega)$ at its output to broaden the bandwidth of the lumped-element impedance inverter at the carrier amplifier. A broadband post-matching network transforms the load resistance into the optimum load impedance required at the common node. A DPA designed using this architecture 42--58\% drain efficiency at 6-dB back-off in 0.9--1.8 GHz (70.7\%).

The continuous-mode theory of power amplifiers was proposed in \cite{cripps09} to provide an extended design space for fundamental and harmonic load impedances. The fundamental load impedance is extended from a resistance $R_{opt}$ to a complex impedance, while the harmonic impedances are modified appropriately (e.g., the short-circuit second-harmonic impedance is replaced with a reactive impedance). This extended design space enables realization of broadband harmonic-tuned power amplifiers. This idea was extended to a DPA in \cite{chen16-3}. Its architecture is similar to the post-matching DPA, where a post-harmonic tuning network has replaced the output matching network. This network is designed to provide optimum load impedance at fundamental and harmonic frequencies. It is shown that using an optimally designed post-harmonic tuning network in the DPA, the average output power and efficiency, for a 20-MHz 7.5-dB PAPR LTE signal, can be improved by 2~dB and 10\%, respectively. A GaN DPA designed based on this architecture features 52--66\% drain efficiency at 6-dB back-off in 1.65--2.75 GHz (50\%). Another DPA using a similar architecture achieves 200 W output power and 40--52\% drain efficiency at 6-dB back-off in 1.7--2.7 GHz (47\%) \cite{chen17-2}. In \cite{shi18}, the output parasitic impedance of the peaking amplifier is employed to provide the optimum load impedance conditions for continuous-mode operation of the carrier transistor. The DPA achieves 46.5--63.5\% drain efficiency at 6-dB back-off in 1.6--2.7 GHz (53\%).

\subsection{Distributed DPA}

\begin{figure}[!t]
\centering
\includegraphics[width=2.8in]{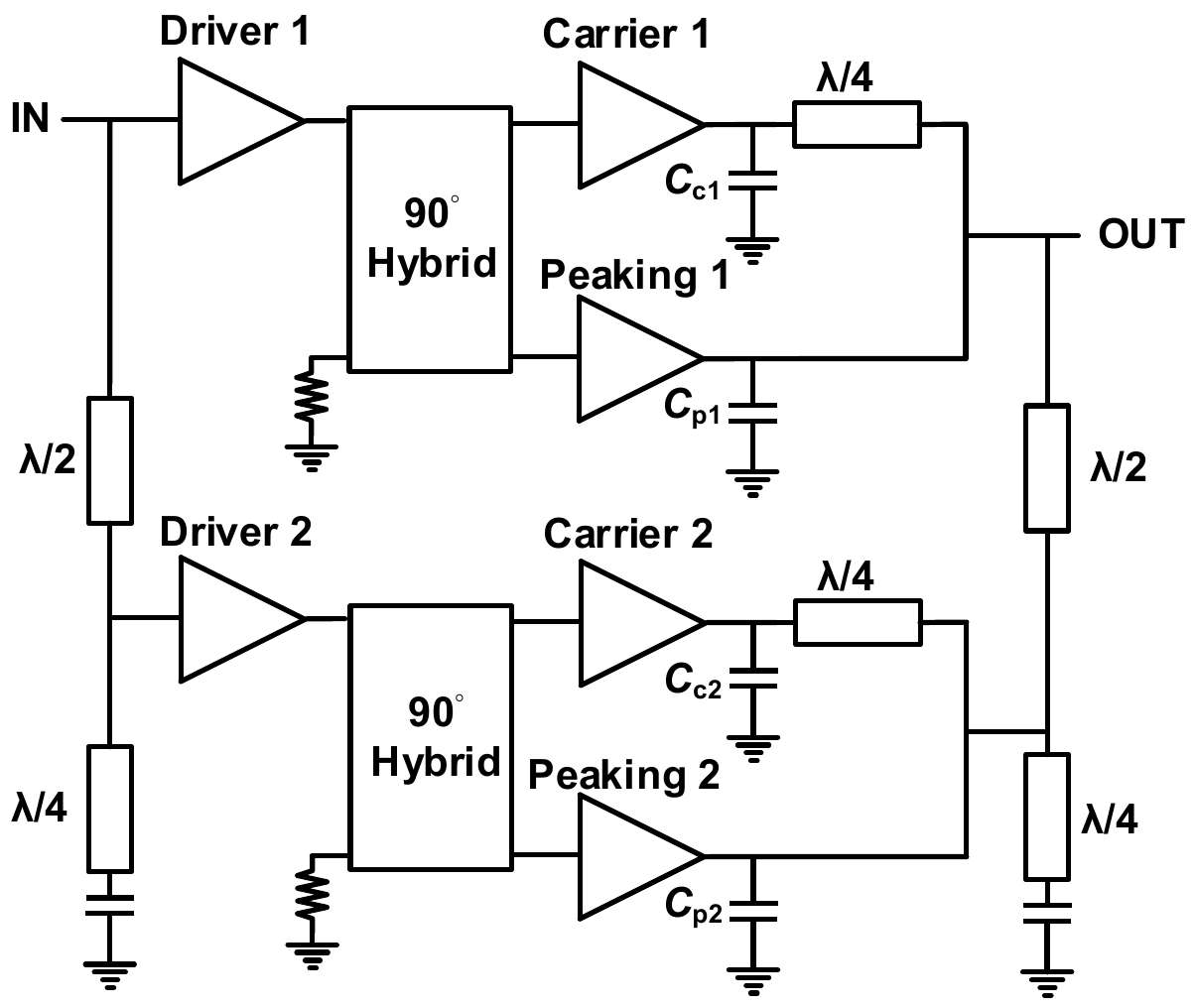}
\caption{The distributed two-way DPA architecture.}
\label{distributed-DPA}
\end{figure}

A distributed amplifier can provide broad bandwidth by absorbing the transistors' input and output parasitic capacitances into the transmission lines connected to the gate and drain \cite{nikandish18}. A broadband distributed DPA architecture was proposed in \cite{lee09}, in which two DPAs with their driver amplifiers are used in the single-ended dual-fed distributed structure without the need for two-way power divider and combiner, shown in Fig.~\ref{distributed-DPA}. 
This architecture inherits some features of the distributed amplifier including absorption of the peaking amplifiers' output capacitance into the output transmission line. However, the output parasitic capacitance of the carrier amplifiers and the impedance inverters still limit the bandwidth.
A GaN DPA reported in \cite{lee09} provides average power-added efficiency (PAE) of 15--25\% in 2.06--2.22 GHz, for a single-carrier WCDMA signal with 10-dB PAPR.

\subsection{Dual-Input DPA}

\begin{figure*}[!t]
\centering
\includegraphics[width=4.0in]{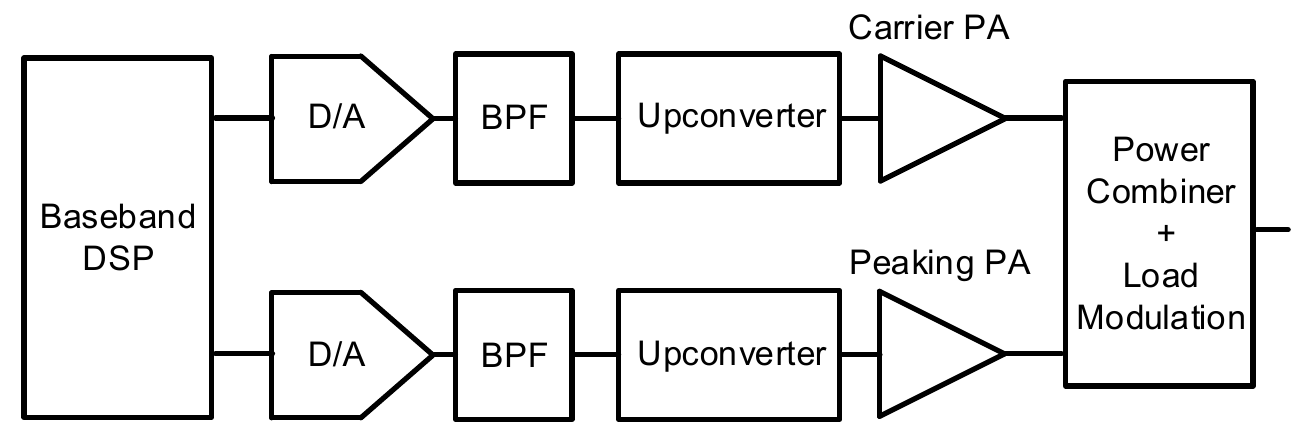}
\caption{Transmitter architecture with dual-input DPA \cite{darraji16}.}
\label{DPA-TX}
\end{figure*}

In the conventional DPA, the two transistors exhibit different drain current profile, gain, and input impedance. Several analog techniques have been proposed to mitigate these issues, including uneven input power splitting, asymmetric DPA architecture, and adaptive gate biasing \cite{darraji16}. A generic approach is to consider the DPA as a dual-input amplifier where magnitude and phase of each input signal can be controlled separately to achieve the optimal operation. The DPA bandwidth can be extended by a frequency-dependent input signal distribution. The efficiency and gain of the DPA can be improved by an adaptive input signal splitting where most of the input power is delivered to the carrier transistor at back-off, while a larger portion is directed into the peaking transistor at peak power \cite{darraji16}, \cite{nick10}, \cite{darraji12}, \cite{andersson13}, \cite{darraji15}. A dual-input DPA can be integrated into a transmitter system to enable digital control of the two input signals with input power and frequency, shown in Fig.~\ref{DPA-TX}. This architecture also facilitates implementation of digital pre-distortion (DPD) algorithms to mitigate the DPA nonlinearity. However, the extra signal processing overhead can limit its application in 5G wireless transmitters with large modulation bandwidths and operation at mm-wave bands.

\begin{figure}[!t]
\centering
\includegraphics[width=3.4in]{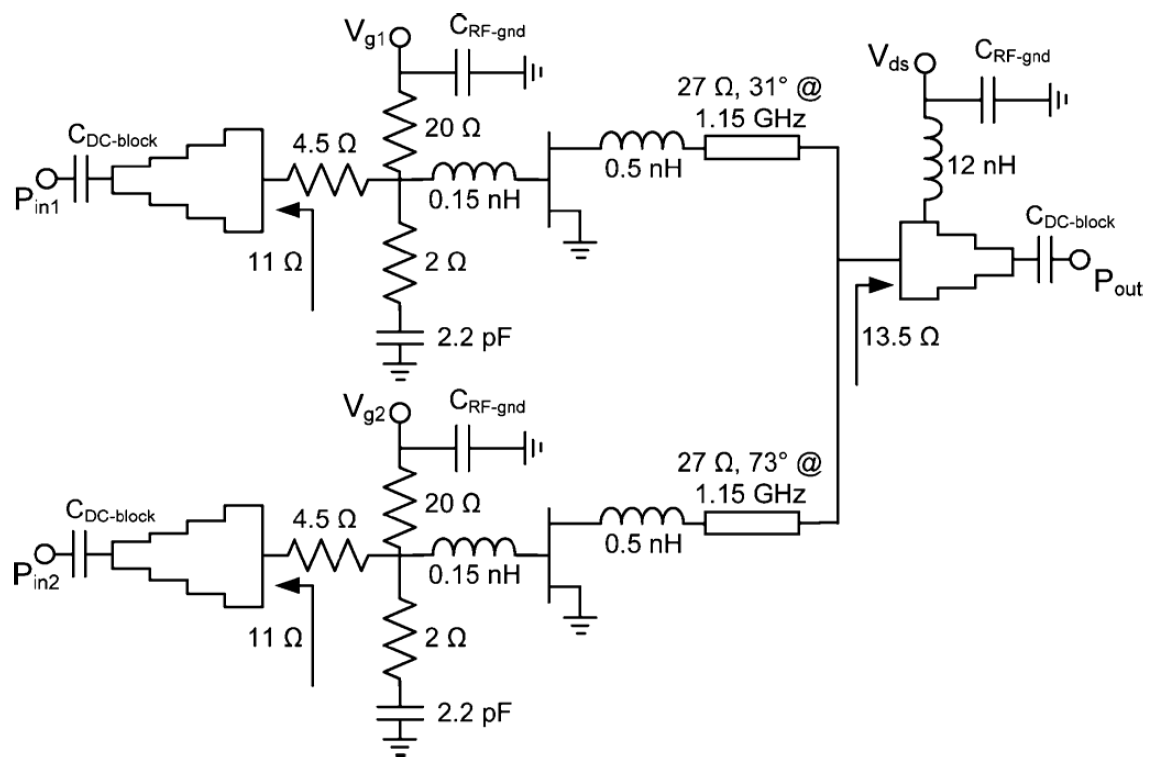}
\caption{The broadband dual-input DPA circuit using stepped impedance transformers reported in \cite{andersson13}.}
\label{doherty-outphasing}
\end{figure}

In \cite{andersson13}, a 1--3 GHz dual-RF input PA based on Doherty-Outphasing continuum was proposed, where relative amplitude and phase of the two input signals are optimally controlled. The DPA circuit designed using GaN HEMTs is shown in Fig.~\ref{doherty-outphasing}. The broad bandwidth is achieved by absorbing parasitic capacitances of the transistors into the matching circuits and further by using stepped impedance transformers. The DPA achieves 45--68\% efficiency at peak power and 48--65\% efficiency at 6-dB back-off in 1--3 GHz (100\%). Almost the same efficiency is obtained at the peak power and back-off as a result of optimum relative amplitude/phase of the two input signals.

\subsection{Transformer-Based Power Combining PA}

\begin{figure}[!t]
\centering
\includegraphics[width=2.2in]{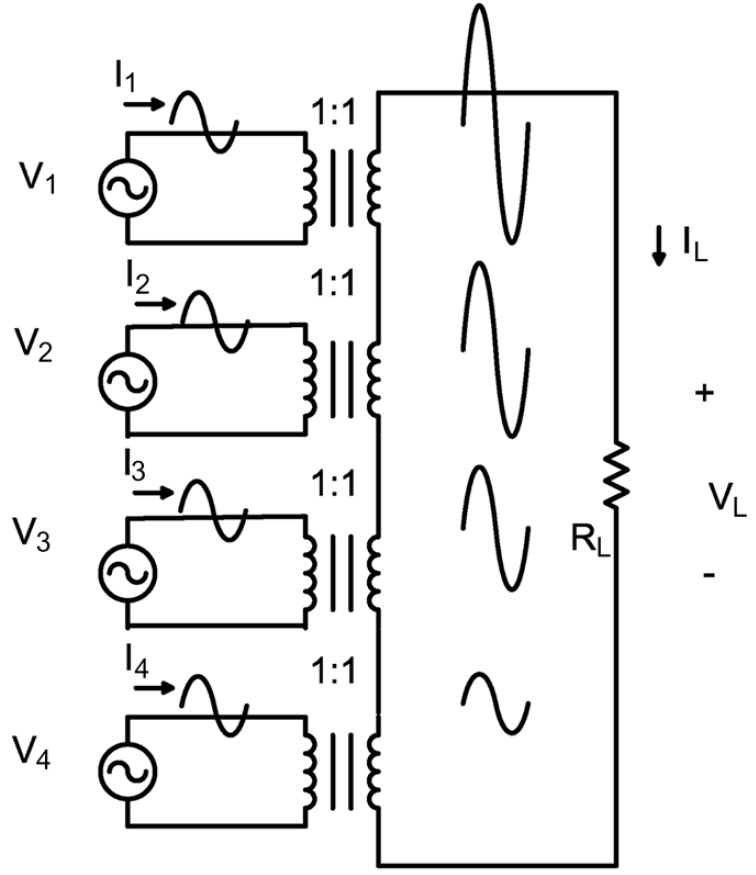}
\caption{Conceptual transformer-based power combining PA \cite{liu08}.}
\label{TR-PA}
\end{figure}

A transformer-based voltage combiner was proposed in \cite{liu08} to combine RF power generated from several low-voltage CMOS amplifiers. The output combiner circuit, shown in Fig.~\ref{TR-PA}, operates as a series voltage adder, where the required output power can be controlled by turning the unit amplifiers on and off. This architecture can modulate the load impedance seen by each unit amplifier. The bandwidth is limited by output parasitic capacitances of the unit amplifiers, switches, and the transformer.

A 2.4-GHz PA implemented in a 0.13-$\mu$m CMOS process achieves 27~dBm peak output power with 32\% drain efficiency at saturation. At 2.5~dB output power back-off, i.e., when one of the four unit amplifiers is turned off, the drain efficiency is 31.5\%, very close to the drain efficiency at peak power.
This PA architecture can provide broad bandwidth by combining small wideband PA cells. Moreover, it can be used to realize reconfigurable transmitters that their output power level can be controlled according to the operation scenario.

\subsection{Transformer-Less Load Modulation PA}

\begin{figure}[!t]
\centering
\includegraphics[width=3.4in]{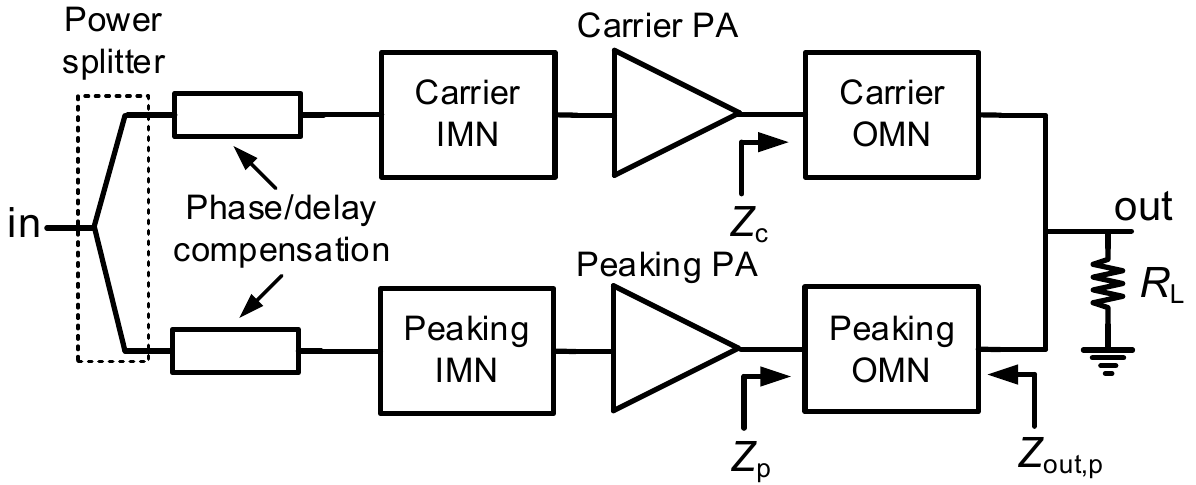}
\caption{Broadband transformer-less load modulation PA architecture \cite{akbarpour12}.}
\label{parallel-tllm}
\end{figure}

A transformer-less load modulation architecture is introduced in \cite{akbarpour12}, \cite{shao14}, \cite{watanabe15} which does not require bandwidth-limiting transmission line impedance transformers or offset lines. The load modulation is performed by broadband output matching networks that transform two load impedances into the optimum values at both peak power and back-off \cite{akbarpour12}.
The PA architecture based on this technique is shown in Fig.~\ref{parallel-tllm}. The carrier amplifier's matching network is designed to present an optimum impedance to the transistor for maximum efficiency at back-off. It also provides a close to optimum impedance at peak power. The peaking amplifier's matching network provides an optimum impedance to the transistor at peak power. It should also exhibit a high output impedance at back-off to prevent power leakage from the carrier amplifier. Finally, the output currents of the two amplifiers should be in-phase at the peak power for proper power combining. This condition can be satisfied with phase alignment networks at the amplifiers inputs. Using this architecture, a GaN PA is designed with 40-45\% drain efficiency at 6-dB back-off in 2.0--2.45 GHz (20\%).
In \cite{watanabe15}, a GaN PA with series-connected load is designed using the similar technique. It achieves 20--50\% drain efficiency at 6-dB back-off in 1.6--2.0 GHz (22\%).

\section{Integrated Circuit Design of DPA}

From the standpoint of 5G user equipment (UE) and small-cell base stations, an IC implementation of the DPA is desired from size/volume reasons. Moreover, with the allocation of mm-wave frequency bands to 5G, IC realization is essential to achieve high performance in the presence of high parasitic components and losses. However, most of the DPAs presented above operate at relatively low frequencies ($<$\,4\,GHz in Table \ref{table_DPAs}) and are implemented as discrete-component circuits where losses and size of passive components are not the most important concerns. The design of broadband IC DPAs is more challenging than that of the conventional broadband high-efficiency PAs \cite{nikandish13}, \cite{nikandish14}, since extra conditions must be satisfied at the peak and back-off output power levels. There are several issues that should be considered in the IC design of broadband DPAs to address the 5G requirements:
\begin{enumerate}
\item Parasitic capacitances can degrade the gain and limit the bandwidth.
\item Losses in transistors and passive components, mainly transmission lines and inductors, degrade the efficiency.
\item Size of quarter-wavelength transmission lines and inductors in the impedance inverter and input power splitter can become excessively large for IC implementations.
\item Current density limitation of the transmission lines impose constraints on their minimum width and hence limit the maximum realizable characteristic impedance.
\item Specific IC design rules, e.g., minimum spacing and density of metal layers, can degraded performance of the DPA by limiting the coupling coefficient of transformers and the quality factor of inductors.
\item Low gain of the class-C biased peaking transistor in high frequencies degrades the DPA gain and efficiency.
\item Mismatch in the gain and phase responses of the carrier and peaking amplifiers, arising from their different bias conditions, makes it challenging to achieve high efficiency at broad bandwidth.
\item In the presence of process variations and mismatches between the two amplifier paths, it is difficult to maintain amplitude and phase linearity in a wide power range, which is an essential requirement for the 5G complex-modulated signals.
\item Large PAPR of 5G signals, e.g., 9.6~dB for 64-QAM with OFDM, requires an asymmetric DPA structure, which as discussed in Section II-A has a larger impedance transformation ratio at back-off and hence narrower bandwidth. Moreover, the peaking transistor would have larger parasitic capacitances that further limit the bandwidth.
\end{enumerate}

These issues render the techniques discussed so far in this paper less effective, if not impractical, for the integrated DPAs and indicate the need for further research. In this section, we present a review of integrated DPAs in RF and mm-wave frequency bands.

\subsection{RF Bands}

\begin{figure*}[!t]
\centering
\includegraphics[width=5.0in]{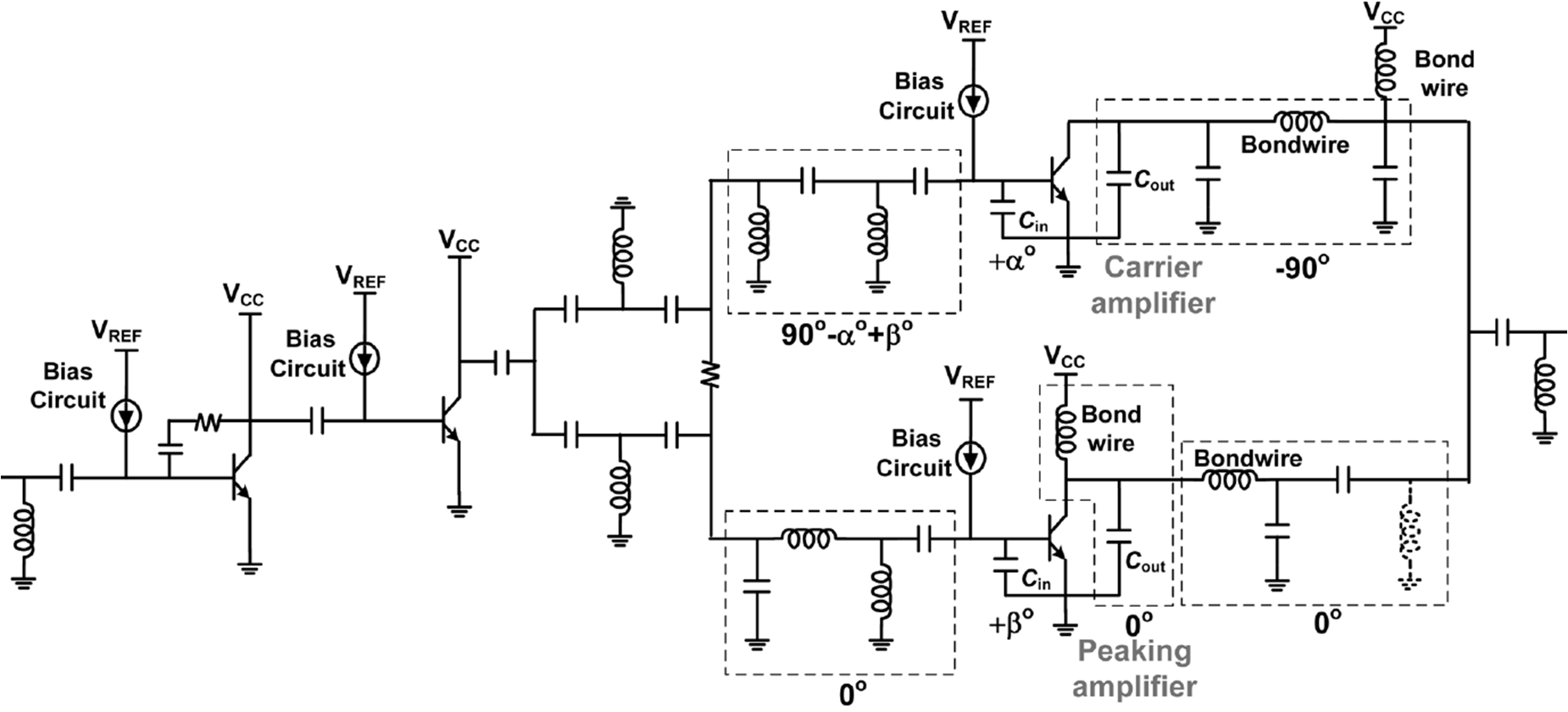}
\caption{Broadband DPA circuit implemented in a GaAs HBT process \cite{kang11}.}
\label{DPA-kang11-2}
\end{figure*}

The sub-6 GHz frequency band, including both unlicensed and various licensed spectra, will be deployed in 5G, especially for delivering fixed wireless services at long distances. The available design techniques for IC DPAs are investigated here to provide insights for further 5G developments. A broadband DPA implemented in a GaAs HBT process \cite{kang11} is shown in Fig.~\ref{DPA-kang11-2}. Output parasitic capacitances of the transistors are absorbed into the lumped-element output network. The inductors are implemented off-chip to lower their losses. For a 10-MHz LTE signal with 7.5~dB PAPR, the DPA delivers an average output power of 27.5~dBm and the PAE of 36\% at 1.85 GHz. The average PAE of over 30\% is obtained in 1.6--2.1 GHz (27\%).
A similar approach was used in \cite{jee15} to design a broadband DPA in a 0.25-$\mu$m GaN-HEMT MMIC process. The output network is designed to provide the same impedance transformation ratio for both the carrier and peaking amplifiers. The DPA achieves average drain efficiency of 46--53\% and average output power of 33.1~dBm in 2.1--2.7 GHz (25\%) for a 10-MHz LTE signal with 7.2~dB PAPR.

\begin{figure*}[!t]
\centering
\includegraphics[width=4.5in]{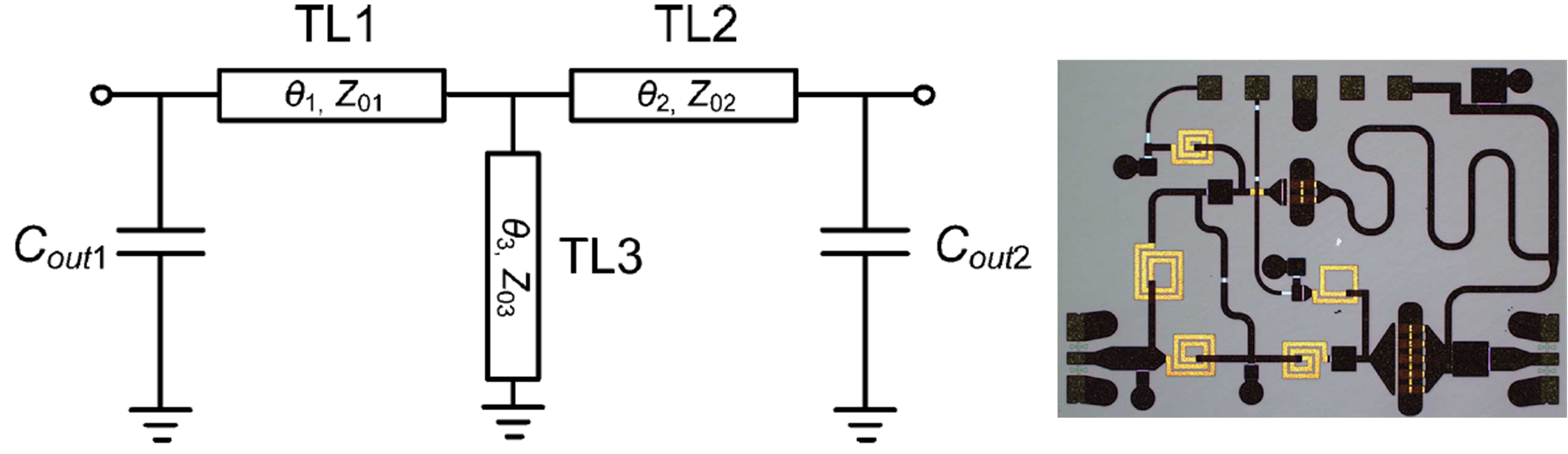}
\caption{The impedance inverter network and 0.25-$\mu$m GaN-HEMT MMIC DPA reported in \cite{gustafsson13-2}.}
\label{DPA-gustafsson13}
\end{figure*}

In \cite{gustafsson13-2}, a compact impedance-inverter network was proposed using a Tee-structure of transmission lines and output parasitic capacitances of the transistors (Fig.~\ref{DPA-gustafsson13}). A DPA is fabricated using a 0.25-$\mu$m GaN-HEMT MMIC process, and achieves the peak output power of 35~dBm, PAE of 38--50\% at peak power and 24--37\% at 9~dB back-off, in 6.8--8.5 GHz (22\%).

\begin{figure*}[!t]
\centering
\includegraphics[width=5.0in]{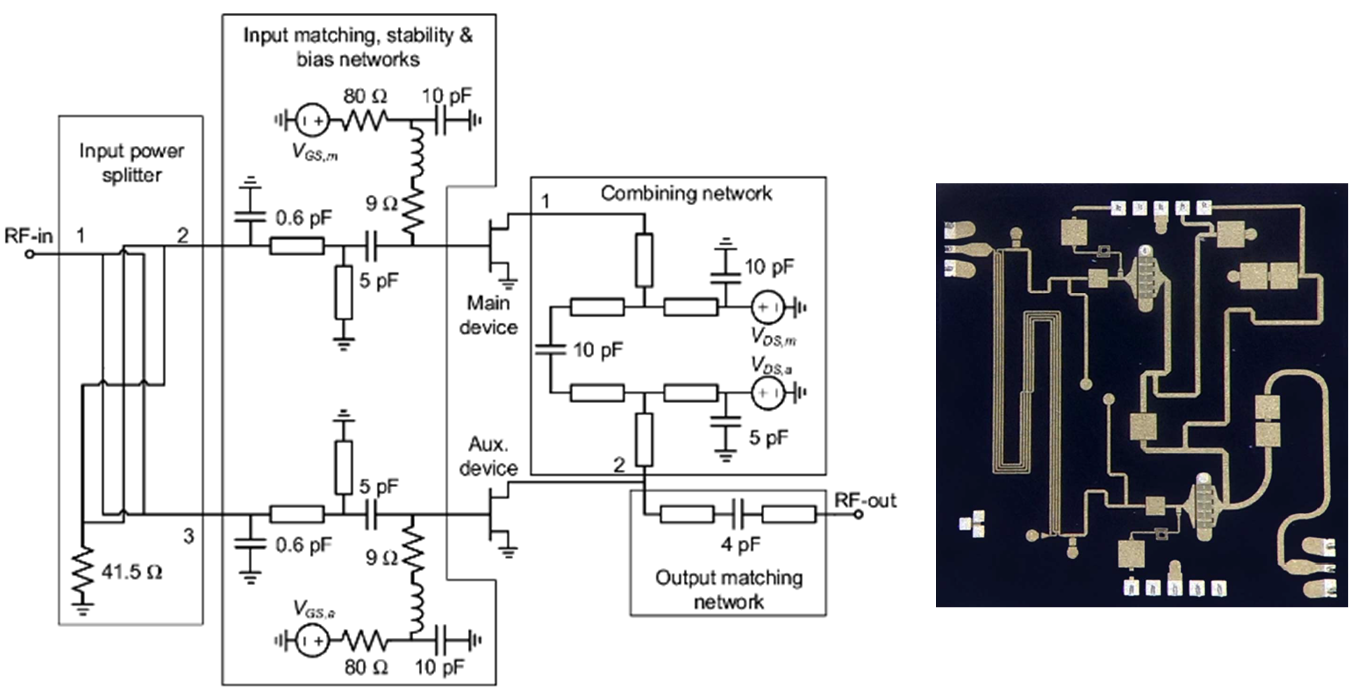}
\caption{The 0.25-$\mu$m GaN-HEMT MMIC DPA reported in \cite{gustafsson14}.}
\label{DPA-gustafsson14}
\end{figure*}

A broadband 0.25-$\mu$m GaN-HEMT MMIC DPA based on the modified load-modulation network is presented in \cite{gustafsson14}. The DPA circuit is shown in Fig.~\ref{DPA-gustafsson14}, where a power combining network absorbs drain-source parasitic capacitances and provides asymmetric drain biases for the transistors. A Lange coupler is used as the input power splitter with broadband quadrature-phase response. The DPA provides 36~dBm peak power and 31--39\% PAE at 9~dB back-off in 5.8--8.8 GHz (41\%).

\subsection{mm-Wave Bands}

With the extensive deployment of mm-wave frequencies envisioned in 5G, new PA design approaches have been developed, e.g., \cite{shakib16}, \cite{vigilante18}, and more research activities are expected in mm-wave DPAs. There are only a few reports of DPAs at mm-wave frequencies to date \cite{agah13}-\cite{indirayanti17}.

One of the key issues in mm-wave DPAs are high losses of the quarter-wave transmission lines which degrade the gain and PAE. In \cite{agah13}, an active phase-shift DPA architecture was proposed where the quarter-wave transmission line at input of the peaking transistor is replaced with a preamplifier. The DPA implemented in a 45-nm SOI CMOS process provides 18~dBm output power, 23\% peak PAE, 17\% PAE at 6~dB back-off, and 7~dB gain at 42 GHz. The bandwidth of this architecture can be extended by using a broadband load for the preamplifier.

\begin{figure*}[!t]
\centering
\includegraphics[width=4.5in]{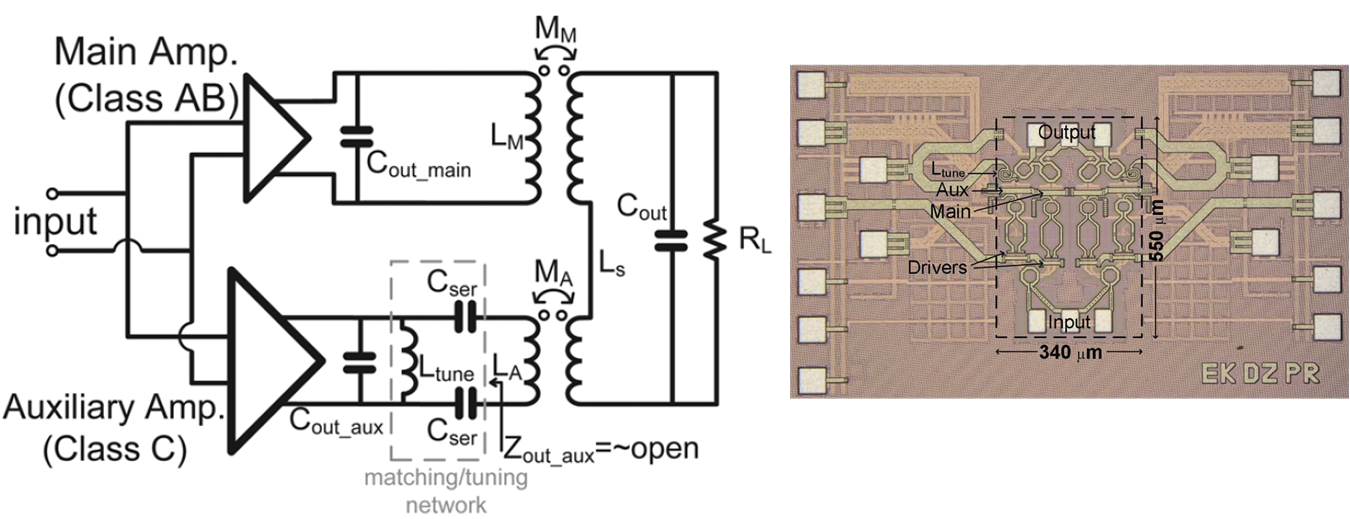}
\caption{Transformer-based DPA implemented in a 40-nm CMOS process \cite{kaymaksut15}.}
\label{DPA-kaymaksut15}
\end{figure*}

A mm-wave transformer-based DPA was proposed in \cite{kaymaksut15}. The circuit is shown in Fig.~\ref{DPA-kaymaksut15}, where an asymmetric series power-combiner with an \textit{LC} tuning circuit at output of the auxiliary amplifier is used to achieve broadband performance. Moreover, each amplifier is implemented as two parallel branches with smaller devices to further extend the bandwidth. A 40-nm CMOS DPA designed using this technique achieves 3-dB bandwidth of 60--81 GHz (30\%). The output power of 20.1~dBm and peak PAE of 11.4\% are maintained over 58--77 GHz. The 6-dB back-off PAE is 7\% at 72 GHz.

\begin{figure}[!t]
\centering
\includegraphics[width=3.3in]{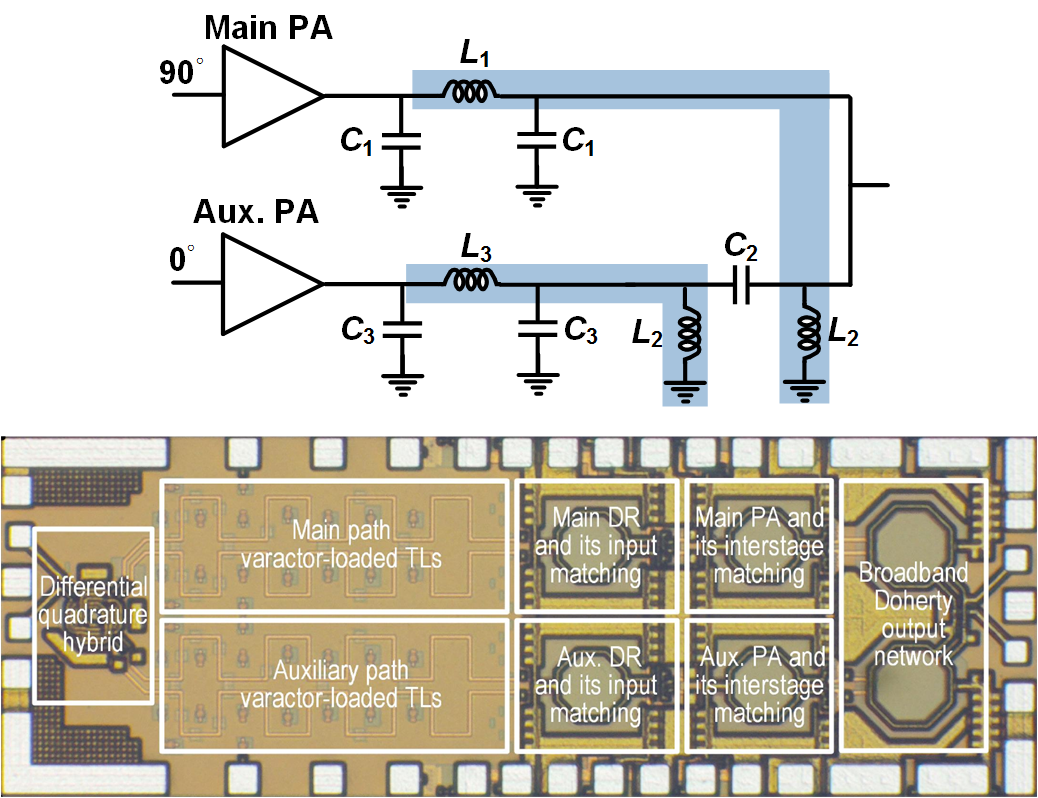}
\caption{The mm-wave multi-band DPA implemented in a 130-nm SiGe BiCMOS process \cite{hu17}.}
\label{DPA-hu17}
\end{figure}

In \cite{hu17}, a two-section peaking network architecture is adopted to design a broadband DPA that covers multiple mm-wave 5G frequency bands.
The transmission lines are replaced by lumped-element circuits, as shown in Fig.~\ref{DPA-hu17}, in which two pairs of coupled inductors, $L_1$-$L_2$ and $L_3$-$L_2$, are realized as transformers. The DPA circuit is composed of differential output and driver amplifier stages, input quadrature hybrid, and varactor-loaded transmission lines to adjust the relative phase shift of the carrier and peaking paths. A power-aware adaptive uneven-feeding scheme is used to gradually deliver more power to the peaking amplifier as the input power increases. Different varactor settings are used for 28, 37, and 39 GHz bands. The DPA, implemented in a 130-nm SiGe BiCMOS process, achieves 3-dB small-signal gain bandwidth of 23.3--39.7 GHz (52\%), and 1-dB saturated output power bandwidth of 28--42 GHz (40\%), collectively in the two settings. The DPA also demonstrates 17~dBm peak output power, 20--23\% peak PAE, and 13--17\% PAE at about 6-dB back-off.

\begin{figure}[!t]
\centering
\includegraphics[width=2.8in]{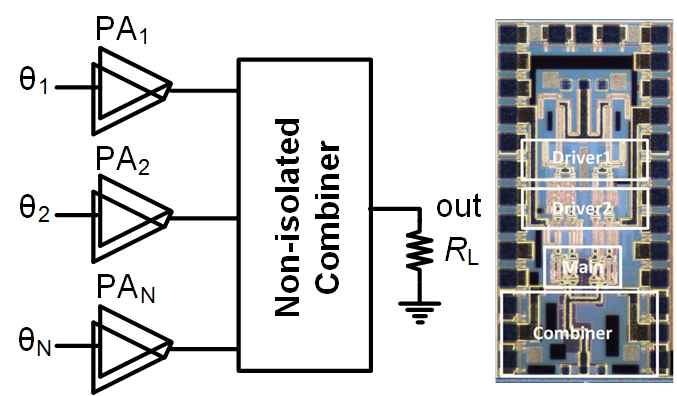}
\caption{The PA architecture with simultaneous reconfigurable frequency and back-off, implemented in a 130-nm SiGe BiCMOS process \cite{chappidi18}.}
\label{reconfig_PA}
\end{figure}

A promising technique for simultaneous frequency and back-off reconfigurability in a mm-wave PA was proposed in \cite{chappidi18}. It is shown that using an asymmetric power combiner that exploits the interactions of PA cells, optimal impedances can be synthesized across the frequency and back-off reconfiguration. Shown in Fig.~\ref{reconfig_PA}, the output power level can be controlled by switching the PA cells, while the impedance presented to the PA cells is dependent on phase of the signals in all paths. Therefore, phase of the input signals can be adjusted to achieve a broadband operation at a given output power level. This architecture can be considered as a generalized combination of the dual-input DPA and transformer-less load modulated PA discussed in Section III. A PA is designed using this technique with two combined paths, each composed of eight PA cells, and adopts an input phase-shift network based on a variable delay line with a varactor bank. The PA, implemented in a 130-nm SiGe BiCMOS process, provides peak output power of 23.7~dBm, peak efficiency of 34.5\%, and 6-dB back-off efficiency of 16--22\% across a broad mm-wave band of 30--55 GHz (62\%).

In summary, only a few of the bandwidth extension techniques developed for the DPA are adopted in IC implementations. The bandwidth and efficiency of the developed IC DPAs are also much lower than their discrete counterparts. While the worse performance partially originates from the limitations of IC processes, new approaches should be developed to effectively use the available bandwidth extension techniques in IC DPAs. For mm-wave bands, where effects of losses and parasitic components are more critical, further research is expected to architect high-performance DPAs.

\section{Conclusion}

The Doherty PA (DPA) is a promising architecture for 5G wireless transmitters that enables efficient amplification of complex-modulated signals with large PAPR. To accommodate the unprecedented increase of data rates and frequency bands envisioned in 5G, the bandwidth of the DPA should be extended. In this paper, we presented a comprehensive review of the DPA bandwidth enhancement techniques and broadband design methodologies published in the literature. Many techniques have been developed for low-frequency discrete DPA circuits and most of them cannot be directly employed in IC implementations. This indicates the need for further research to develop broadband design techniques that address challenges and limitations of IC processes. From the techniques investigated in Section III, the DPA with modified impedances of transmission lines is a feasible architecture for IC implementation. The two-section peaking network needs three quarter-wavelength transmission lines that normally take excessive on-chip area. However, as shown in Section IV, it is possible to derive equivalent lumped-element circuits with a compact IC realization. The DPA with short-circuited stub can be implemented on chip and can absorb parasitic capacitance of the peaking transistor into the stub circuit. The dual-input DPA architecture is useful in transmitter systems where signal modulation, pre-distortion, and conditioning can be performed in digital domain and applied to the DPA to improve its performance. The transformer-based power combining PA, originally developed for IC PAs, can be effectively used at mm-wave frequencies. The transformer-less load modulation of PA can also be adopted in IC DPAs, as the large impedance inverters can be replaced by lumped-element circuits.
It is expected that more research activities will be attracted to the design of integrated DPA circuits in the future, specially at mm-wave frequencies, in order to address the requirements of 5G wireless transmitters.

\section*{Acknowledgment}
The authors would like to thank the late Prof. Thomas J. Brazil for helpful comments on the manuscript.
This research has received funding from the European Union's Horizon 2020 Research and Innovation Program under the Marie Sklodowska-Curie grant agreement number 713567, and Science Foundation Ireland (SFI) under grant numbers 13/RC/2077 and 16/IA/4449.


\end{document}